\documentclass{aa}
\usepackage{graphicx}
\usepackage{txfonts}
\usepackage{tikz}
\usetikzlibrary{shapes.geometric, arrows,positioning}

\usepackage{subfig}
\titlerunning{The edges of galaxies}

\begin{document}

   \title{The edges of galaxies: tracing the limits of star formation}
   \author{Nushkia Chamba,
          \inst{1}\thanks{email: nushkia.chamba@astro.su.se}
          Ignacio Trujillo
          \inst{2, 3}
          \and
          Johan H. Knapen
          \inst{2, 3}
          }
   \institute{The Oskar Klein Centre, Department of Astronomy, Stockholm University, AlbaNova, SE-10691 Stockholm, Sweden
        \and 
            Instituto de Astrof\'isica de Canarias (IAC), E-38205 La Laguna, Tenerife, Spain
         \and
             Departmento de Astrof\'isica, Universidad de La Laguna (ULL), E-38200 La Laguna, Tenerife, Spain
             }

   \date{Received YYY; Accepted XXX.}


\abstract{The outskirts of galaxies have been studied from multiple perspectives for the past few decades. However, it is still unknown if all galaxies have clear-cut edges like everyday objects. We address this question by developing physically motivated criteria to define the edges of galaxies. Based on the gas density threshold required for star formation, \textit{we define the edge of a galaxy as the outermost radial location associated with a significant drop in either past or ongoing in-situ star formation}. We explore $\sim$1000 low-inclination galaxies with a wide range in morphology (dwarfs to ellipticals) and stellar mass ($10^7 M_{\odot} < M_{\star} < 10^{12}M_{\odot}$). The location of the edges of these galaxies ($R_{\rm edge}$) are visually identified as the outermost cut-off or truncation in their radial profiles using deep multi-band optical imaging from the IAC Stripe82 Legacy Project. We find this characteristic feature at the following mean stellar mass density which varies with galaxy morphology: $2.9\pm0.10\,M_{\odot}$/pc$^2$ for ellipticals, $1.1\pm0.04\,M_{\odot}/$pc$^2$ for spirals and $0.6\pm0.03\,M_{\odot}/$pc$^2$ for present-day star forming dwarfs. Additionally, we find that {$R_{\rm edge}$} depends on its age (colour) where bluer galaxies have larger $R_{\rm edge}$ at a fixed stellar mass. The resulting stellar mass--size plane using $R_{\rm edge}$ as a physically motivated galaxy size measure has a very narrow intrinsic scatter ($\lesssim 0.06$\,dex). These results highlight the importance of new deep imaging surveys to explore the growth of galaxies and trace the limits of star formation in their outskirts.}
 

\maketitle

\section{Introduction}

Galaxies grow and evolve through two main channels: \textit{in-situ} star formation via the conversion of gas into stars and \textit{ex-situ} stellar and gas accretion via merging and interactive events with its neighbourhood \citep[e.g.][]{1972toomre, 1978white, 1983efstathiou}.  While merging events are expected to happen predominantly in the most massive galaxies, current stellar mass growth via in-situ star formation occurs in the majority of dwarfs and spiral galaxies, and depends on the density of gas in these systems. In other words, stars can form in a galaxy as long as the density of gas surpasses a critical threshold \citep[e.g.][]{1972quirk, 1980fall, 1989kennicutt, 2004schaye}. The radial location of the `edge’ of star formation as defined by this critical gas density threshold in a galaxy is thus a physically meaningful way to study the growth and evolution of galaxies. \par 

This idea has recently been proposed by \citet{2020tck} and \citet{2020ctk} as a new, physically motivated definition of galaxy size \citep[see also][for a historical review on galaxy size measures]{2020sizeschamba}. To make this definition operative, in these studies we specifically chose a stellar mass density of 1\,$M_{\odot}/$pc$^2$ as a proxy to locate the gas density threshold required for star formation in galaxies based on theoretical \citep[e.g.][]{2004schaye} and observational \citep[e.g.][]{2019cristina} evidence. This statement is based on the abrupt drop in the ultra-violet radial profiles of two Milky Way-like galaxies reported by \citet{2019cristina} in the region where their stellar mass density profiles are truncated (at 1\,$M_{\odot}/$pc$^2$ after correcting for inclination). Consequently, the radial location of this isomass contour was used as a size measure (dubbed $R_1$), uniquely associated to the visual location of the edges of galaxies. \citet{2020tck} showed that the resulting size--stellar mass relation over five orders of magnitude in stellar mass ($10^7 M_{\odot} < M_{\star} < 10^{12}M_{\odot}$) produced an extremely tight distribution, with an intrinsic scatter (i.e. 0.06 dex) three times smaller than when using \citet{1948dev} effective radius, a popular measure for galaxy size defined as the radius that encloses half the total light of a galaxy. Furthermore, \citet{2020ctk} showed that in contrast to the effective radius that depends on the concentration of light in galaxies, $R_1$ is a much better representation of the boundaries of galaxies to fairly compare the sizes of distinctive galaxy populations or morphologies whose light distributions are very different. \par 

While the above results are very promising, the truncation at 1\,$M_{\odot}/$pc$^2$ has thus far only been observed at the edges of Milky Way-like disk galaxies \citep{2019cristina}. As the exact stellar mass surface density depends on the efficiency of transforming gas into stars, the fixed value at 1\,$M_{\odot}/$pc$^2$ cannot be assumed to hold also for galaxies of other morphologies and/or stellar mass. For this reason, we seek to measure the star formation threshold and consequently the sizes of galaxies belonging to wide ranges in morphology, from dwarfs to elliptical galaxies, and stellar mass. As a proxy for our measurement, we search for a change in slope, cut-off or truncation in the radial stellar mass density profiles of the galaxies in our sample. The origin of truncations in the outer profiles of galaxies is an open question and several interesting scenarios have been proposed on its connection with the evolution of galactic disks \citep[see][for a review]{2011vanderkruit}. However, there is growing evidence that truncations are intimately linked to thresholds in star formation activity \citep{1989kennicutt, 2008roskar, 2017elmegreen, 2019cristina}. Therefore, searching for a truncated feature in the stellar mass density profiles of different types of galaxies is also a step towards addressing the origin of this signature. \par 
This paper is organised as follows. We explain the meaning and concept of galaxy edges in broader terms in Section \ref{sect:concept}. The imaging data and sample selection is described in Section  \ref{sect:sample}. The details of the methods used can be found in Sections \ref{sect:methods} and \ref{sect:define_edge}. The results are shown in Section \ref{sect:results} and discussed in Section \ref{sect:discuss}. Our main conclusions are presented in Section \ref{sect:conclusions}. We assume a standard $\Lambda$CDM cosmology with $\Omega_m$=0.3, $\Omega_\Lambda$=0.7 and H$_0$=70 km s$^{-1}$ Mpc$^{-1}$.

\section{An intuitive and physically motivated definition of the edge of a galaxy}
\label{sect:concept}

In computer vision, the edge of an everyday object is detected where there is a sharp contrast or change in their properties such as brightness, colour, shape or texture \citep[see the recent review by][]{2022jing}. The location of these features are frequently used to define the object's size. Automatically segmenting or detecting the edges of light sources such as galaxies using astronomical images, however, is a nontrivial task \citep[e.g.][]{2021haigh}. But this issue can be addressed by developing physically motivated criteria and features to define the edges of galaxies. In this paper, \textit{we define the edge at the outermost radial location associated with a  significant drop in either ongoing or past in-situ star formation}. Consequently, the light beyond the edge is mostly contributed by ex-situ stars belonging to the stellar halo (\citet{2020tck, 2020font, 2022song} and see \citet[][for a clear example]{2021trujillo}). The above reasons make our definition of the edge of a galaxy intuitive to the broader concept of the edge of an object, motivate its use as physical measure to fairly represent and compare the sizes of all galaxies, and as a method to define the outer stellar halo \citep[see][]{2020tck, 2020ctk, 2020sizeschamba}. \par 

But how can one detect the edges of galaxies and where are they located? As discussed in the Introduction, there is growing evidence to suggest that truncations, a change in slope or cut-off feature in the outskirts of a galaxy's radial profile, are indicative of a star formation threshold, i.e. a drop in \textit{in-situ} star formation \citep{1989kennicutt, 2008roskar, 2017elmegreen, 2019cristina, 2022simon}. For this reason, in this work we use the truncation as a signature of the edge. However, we prefer the term `edge' over the classical `truncation' in this paper because truncations were specifically characterised for edge-on Milky Way-like galaxies by \citet{1979vanderkruit} and \citet{1981vandekruita, 1981vandekruitb} while we aim to study these features in low-inclination galaxies. We select low-inclination  galaxies because there are only a few studies in the literature where their outskirts have been explored \citep[e.g][]{2011hunter, 2017peters, 2022watkins} and in smaller sample sizes than what we examine here ($\sim$ 1000 galaxies). Low-inclination galaxies are also less affected by scattered light due to the point spread function (PSF) \citep[e.g.][]{2001trujillo} compared to edge-on galaxies which makes these galaxies practically advantageous when studying the properties of their edges. \par

For clarity, we point out that our definition of the edge does not depend on \textit{when} star formation occurred in the galaxy, whether it is ongoing or recent as in star forming galaxies, or happened in the distant past as in elliptical galaxies. We are thus capable of implementing our definition on galaxies that have varied evolutionary pathways and comparing them on equal footing. In short, we characterise the edges of different types of galaxies, from dwarfs to giants (Sect. \ref{sect:sample}), within a common physically motivated framework, i.e. edges indicative of a current or past star formation threshold. And we study these edges as a function of galaxy morphology and stellar mass. \par 

The criteria we use to identify edges for each morphological type is detailed in Sect. \ref{sect:define_edge}. As we explain in Sect. \ref{sect:define_edge}, we perform this task using a large variety of evidence, including the stellar mass density profile, colour radial profile and the multi-band optical images. The radial location we use as a signature of the edge will be called $R_{\rm edge}$. Consequently, we use $R_{\rm edge}$ as a physically motivated measure of galaxy size. We determine the stellar mass density at that location ($\Sigma_{\star}(R_{\rm edge})$) and study the resulting size and stellar mass density as a function of galaxy stellar mass and morphology, all at low redshift (Sect. \ref{sect:results}). We then discuss the implications of our results on the formation and evolution of galaxies (Sect. \ref{sect:discuss}). \par

\section{Data and sample selection}
\label{sect:sample}

\subsection{Deep Stripe 82 Imaging}
\label{sect:s82_data}

A deep and wide multi-band survey with sub-kpc spatial resolution is necessary to resolve the edges of a large sample of low-inclination galaxies where mass densities are low and the scale of the feature is of the order of $\sim 1$\,kpc.
The deep $g$- and $r$-band images of the IAC Stripe 82 Legacy Project\footnote{\protect\url{http://research.iac.es/proyecto/stripe82/}} \citep{2016s82legacy, 2018s82rectified} is thus chosen for this work. The dataset is a co-added version of the Sloan Digital Sky Survey \citep[SDSS;][]{2000york} `Stripe 82' \citep{2008jiang, 2009abazajian} that has been optimized for low surface brightness astronomy.  The limiting depth in surface brightness of these images are $\mu_g = 29.1$\,mag/arcsec$^2$ and $\mu_r = 28.5$\,mag/arcsec$^2$, both measured as a 3$\sigma$ fluctuation with respect to the background of the image in $10\times10\,$arcsec$^2$ boxes. Assuming a $\sim 1$\,arcsec spatial resolution for SDSS imaging, we are capable of resolving structures down to $\sim 600$\,pc at the median redshift of $z \sim 0.03$ in our sample. We also make use of the publicly available extended ($R \sim 8$\,arcmin) point spread functions (PSF) of the SDSS survey \citep{2020raul}. \par

We use the same sample of galaxies studied in \citet{2020tck}, namely elliptical (c0-E+ TType) and spiral (S0/a-Im TType) galaxies from \citet{2010nair} and a low-mass sample of (dwarf) galaxies from \citet{2013maraston} within the Stripe 82 footprint. This can be considered the parent sample in our analysis. Galaxies with contaminated outskirts due to very bright stars, Galactic cirrus structures or nearby companion/interacting objects were removed from the initially selected sample. The final parent sample consists of  1005 galaxies (279 ellipticals, 464 spirals and 262 dwarfs) with stellar masses between $10^7M_{\odot}<M_{\star}<10^{12}M_{\odot}$ and redshift $0.01 < z < 0.1$. See \citet{2020tck} for more details. \par 

The sample of late-type galaxies studied in \citet[][hereafter B12]{2012bakos} and \citet[][hereafter P17]{2017peters} is also included in this work (24 galaxies) to complement our investigation in two ways: 1) the galaxies from B12 are located at lower distances and are therefore at a higher spatial resolution: the median distance of galaxies in this sub-sample is 54.2\,Mpc which corresponds to a spatial resolution of about 260\,pc/arcsec. This implies that the edge (if any) should be more prominent for these galaxies, 2) the sample from P17 is interesting because we can study galaxies with low inclinations. Upon examination, three galaxies from the P17 sample, namely UGC 2319, UGC 2418 and NGC 7716 were removed due to heavy scattered light contamination from nearby bright stars.  \par

In order to explore the effect of the inclination on the location of the truncation, we also select an edge-on galaxy from \citet{2018shinnjongho}, UGC09138, with similar rotational velocity after inclination correction than other low-inclination galaxies in our sample (SDSS J001431.85-004415.26, NGC1090, SDSS J011050.82+001153.36, SDSS J031133.38-004434.50). These galaxies have rotational velocities  corrected for inclination \footnote{Values obtained from the HyperLeda database \citep{2014makarov}: \protect\url{http://leda.univ-lyon1.fr/}} between $135\text{km/s} < V_{\textit{rot}} < 155\,$km/s. UGC09138 is outside the footprint of Stripe82 and therefore imaging in the $g$ and $r$ bands from the Sloan Digital Sky Survey (SDSS) DR12 were obtained using the SDSS mosaic tool\footnote{\protect\url{http://dr12.sdss.org/mosaics/}}. Using SDSS images for this galaxy does not affect our analysis. Therefore, a total of 1027 galaxies was analysed in this work. \par

\subsection{GALEX}

As the near and far ultra-violet (NUV and FUV, respectively) imaging of galaxies from \textit{Galaxy Evolution Explorer} \citep[\textit{GALEX};][]{2005galex, 2007morrissey} are sensitive indicators of star formation, this data is an ideal way of tracing the connection between edges and a star formation threshold. While \textit{GALEX} imaging is also well-suited for the goals of this work in terms of depth (the images used here have a surface brightness limit of 29.6$\pm0.5$ mag/arcsec$^2$ (3$\sigma$, 100 arcsec$^2$)), its lower spatial resolution (FWHM$\sim$4.5 and 5.4 arcsec in the FUV and NUV respectively) makes it unfeasible to explore in detail the edges in our full sample. We thus only use \textit{GALEX} imaging for a handful of nearby galaxies in our sample to illustrate our definition of the edge of a galaxy and how it can be located in low-inclination galaxies compared to edge-on configurations using optical data.

We retrieve intensity maps from the Guest Investigator Program (GI) and Medium Imaging Survey (MIS) with the longest exposure times for the same galaxies we selected to explore the effect of the inclination on the location of the edges (see Sect. \ref{sect:s82_data}). This includes edge-on galaxy  UGC09138 (GI4-016007 and GI6-026001), intermediately inclined galaxies SDSS J001431.85-004415.26 (MISGCSN-29100-0389), NGC1090 (MISGCSS-18291-0409o), SDSS J011050.82+001153.36 (GI6-060007 and MISWZS01-30939-0269) and a face-on galaxy SDSS J031133.38-004434.50 (MISGCSS-18648-0410o). We followed \citet{2007morrissey} to convert the intensity to AB magnitudes. \par \bigskip

\section{Methods}
\label{sect:methods}

To derive accurate surface brightness profiles of galaxies, the first steps are to find the elliptical parameters (centre, axis-ratio and position angle) that best describe the galaxy outskirts and to estimate the background of the image. There are two main complications that make these procedureschallenging. One is the masking of all sources in the vicinity of the galaxy in the image and the second is accounting for contamination from scattered light. All these image processing techniques and the measurement of the ellipticity of our sample of galaxies have been previously presented in \citet{2020tck} and summarised in \citet{2020ctk}. The main steps to derive the radial surface brightness, colour and stellar mass density profiles of galaxies are described below:

\begin{itemize}
    \item[$\bullet$] Individual image stamps centred on each galaxy were created in the $g$ and $r$--band with dimensions $600\times600$\,kpc$^2$ for the massive galaxies and $100\times100$\,kpc$^2$ for the dwarfs in their rest frame. The dimensions of these stamps are at least five times greater than the rest-frame sizes of galaxies in our sample which are important for an accurate background subtraction and masking.
    
    \item[$\bullet$] Scattered light due to point sources in each image was removed using the PSFs developed by \citet{2020raul} for the SDSS survey. We used GAIA DR 1 \citep{2016gaia} to initialise the locations and brightnesses of stars. The normalised PSFs were scaled to match the brightnesses of stars with the GAIA filter G < 17 mag at their locations on the image using IMFIT  \citep{2015imfit}.
    
    \item[$\bullet$] All other sources surrounding the galaxy of interest were masked using an automated source detection tool `Max-Tree Objects' (MTO) \citep{2016mto}. We used the version developed in \citet{2021haigh} who demonstrated the algorithm's capabilities in detecting low surface brightness light compared to other tools in the literature.
    
    \item[$\bullet$] Background values for each galaxy were estimated using the fully masked images. We selected all the pixels that remained undetected (i.e. without source light) from the MTO segmentation map and used those regions to compute the mean background value and the associated dispersion. The mean background value was then subtracted from the masked images. The background subtracted images are the ones used to derive radial profiles of galaxies.
     
    \item[$\bullet$] The centre, axis-ratio and position angle of each galaxy was then computed at the location of the 26 mag/arcsec$^2$ isophote in the $g$-band by fitting an ellipse to the spatial distribution of the pixels at this isophote. This isophote is close to a traditional definition of galaxy extension by \citet{1958holmberg} and its location provides an initial estimate of the global shape and size of the galaxy. We visually checked these parameters to ensure the ellipse follows the global shape of the galaxy in its outskirts.

    \item[$\bullet$] We fixed these elliptical parameters and used them to derive the radial surface brightness profiles in $\mu_g$ and $\mu_r$. Flux was averaged over concentric elliptical annuli from the centre of the galaxy to 200 arcsec, which is well beyond the visual extent of the galaxies in our sample. In this way, we were able to verify that our background subtraction was performed accurately \citep[see Sect. 5.3 in][for details]{2020tck}.
    
    \item[$\bullet$] The profiles of the spiral and dwarf galaxies are corrected for the inclination effect following the model developed in \citet[][see their Sect. 5.2]{2020tck}. And all profiles are corrected for redshift dimming as well as Galactic extinction, using the position of the galaxies in the sky as input to NED's calculator\footnote{\protect\url{https://ned.ipac.caltech.edu/ forms/calculator.html}}. The corrected $\mu_g$ and $\mu_r$ profiles are then used to compute the $g-r$ colour and stellar mass density $\Sigma_{\star}$ profiles.
    
    \item[$\bullet$] The stellar mass density profile was computed using the mass-to-light ratio ($M/L$) versus colour relation prescribed by \citet{2015roediger}. Explicitly, for a given wavelength $\lambda$, the relevant equations are:

\begin{align}
\label{eq:mass}
    \log{\Sigma_{\star, \lambda}} = 0.4 (m_{abs,\odot, \lambda}& - \mu_{\lambda}) + \log{(M/L)}_{\lambda} + 8.629 \\
    \log{(M/L)_{\lambda}} =& m_{\lambda} \times (\text{colour}) + b_{\lambda}
\end{align}
{In this work, we use $\lambda=g$ as it is our deepest imaging dataset and the $g-r$ colour to calculate $(M/L)_g$, using $m_{g}=2.029$ and  $b_g=-0.984$ \citep[see Table A1 in][]{2015roediger}.}

\end{itemize}

We visually examine the derived surface brightness and stellar mass density profiles of the galaxies for their edge (i.e. a change in slope or cut-off in their radial profiles as discussed in the Introduction, Sect. \ref{sect:concept} and detailed criteria are specified later in Sect. \ref{sect:define_edge}). We select the radial location of this feature {($R_{\rm edge}$)} and determine the stellar mass density in that location ($\Sigma_{\star}(R_{\rm edge})$). We also measure the colour at the edge location using the $g-r$ profile of the galaxy. From this colour, we determine the age of the stellar population using the extended MILES library in the SDSS bands, assuming the Kroupa Universal IMF\footnote{\protect\url{http://research.iac.es/proyecto/miles/pages/photometric-predictions-based-on-e-miles-seds.php}} \citep{2012emiles}. We use the MILES predictions for a metallicity $[M/H] = 0$ and -0.71 \citep[e.g.][]{2014radburn-smith}. Low metallicities have been observed in the outskirts of galaxies with the wide stellar mass range we study here (i.e. $10^7\,M_{\odot} < M_{\star} < 10^{12}\,M_{\odot}$), from low mass spirals to massive ellipticals \citep[see e.g.][]{2021neumann}. Additionally, as reviewed in \citet{2017elmegreen} and \citet{2017denija}, it is well established that the metal-poor stellar populations of local dwarf galaxies are very similar to the outskirts of disk galaxies. Therefore, both the fixed metallicity values we use from MILES are well-motivated to estimate the age at $R_{\rm edge}$ \citep[see also the recent work by][]{2022cardona}.  \par 
Finally, we quantify the uncertainties in $R_{\rm edge}$ and $\Sigma_{\star}(R_{\rm edge})$ from the main sources: 1) background estimation and subtraction 2) colour to stellar mass conversion and 3) the visual identification of the edge. We follow a similar approach to that described by \citet{2020tck} for the treatment of the first two. Namely, we fix the location of the edge and then follow how the radial profiles move by a random quantity prescribed by the dispersion in the measured background value per image and stellar mass estimate in our procedure. In other words, we follow how the inferred $\Sigma_{\star}(R_{\rm edge})$ changes due to our background and mass estimate if we fix $R_{\rm edge}$ and vice versa. The dispersion in the stellar mass comes from comparing the estimate from  Eqs. \ref{eq:mass} and those published by \cite{2013maraston}. Details of this comparison is provided in the Appendix in \citet{2020tck} and our uncertainty estimations for this work are provided in Sect. \ref{sect:results}. To infer the third source of uncertainty, we used repeated identifications from our visualisation procedure and evaluated the dispersion in these measurements. We discuss the visualisation procedure in more detail in Sect. \ref{sect:visual_procedure}.

In this work, we do not attempt to correct the surface brightness profiles of the galaxies due to the effect of the PSF as the radial location of the edge remains unchanged \citep[see for e.g.][]{2016trujillo}. However, we point out that (in general) the PSF can affect the estimated stellar mass density at the location of the edge. When corrected for the PSF effect, the brightness of either the $\mu_g$ or $\mu_r$ radial profiles would decrease \citep[this effect is clearly visible in][as the galaxy explored in that paper is highly inclined]{2016trujillo}. Consequently, this effect means that all of our $\Sigma_{\star}(R_{\rm edge})$ estimations are upper limits. Considering the low inclinations of the galaxies in our sample, however, we expect that the effect of the PSF will be very mild \citep[see e.g.][]{2001trujillo}. In the case of the IAC Stripe 82 survey, the shape of the PSF in the $g$ and $r$ band filters are very similar \citep[see][]{2020raul}. This statement is also true for GALEX \citep[see Figs. 9 and 10 in][]{2007morrissey}. Given that the stellar mass density is a function of $\mu_g$ and the $g-r$ colour (Eq. 1 and 2), at least at first order, the effect of the PSF can be neglected in the $g-r$ profile. We leave a more detailed analysis on the full effect of the PSF for future work.

\section{Locating the edge of a galaxy}
\label{sect:define_edge}

This section details the procedure and criteria adopted in this work to locate the edges of galaxies (Sect. \ref{sect:visual_procedure}). We then further discuss how the manifestation of the edge in the profile {is physically motivated based on our edge definition and criteria for each morphological type. The criteria used to identify the signature of the edge is discussed for late-type, spiral galaxies (Sect. \ref{sect:orient}), early-type, elliptical galaxies (Sect. \ref{sect:edges_ellipses}) and dwarf galaxies (Sect. \ref{sect:edges_dwarfs}). Several examples of the edges identified in our sample for each galaxy type are also shown.

\subsection{Visualisation procedure and criteria}
\label{sect:visual_procedure}

The visualisation procedure of each galaxy and its profiles is illustrated as a flowchart in Fig. \ref{fig:flowchart}. As motivated in Sect. \ref{sect:concept}, the main criterion we use as a signature of the edge is the change in slope in the outermost region of the galaxy's radial profiles. For each galaxy, we first examine their stellar mass density profile $\Sigma_{\star}(R)$ for {the edge} and mark the radial location $R_{\rm edge}$. If we are unable to locate a signature for edge, we proceed to examine the surface brightness profiles in $g$ and $r$, followed by the $g-r$ colour profile and do the same. \par 

In the case of the colour profile, the criteria we use as a signature of the edge depends on the morphology of the galaxy. For disk galaxies, we search for the location of a sudden reddening in the outer part of the profile, indicative of the end of the star forming disk (Sect. \ref{sect:orient}). For elliptical galaxies, we use the location of a sudden transition towards bluer $g-r$ colours in the profile, potentially related to an outer envelope which assembled more recently compared to the galaxy's (redder) central regions (Sect. \ref{sect:edges_ellipses}). And for dwarf galaxies, the signature of the edge in the colour profile may appear either as a transition to bluer or redder outskirts, a reflection of the varied star formation histories possible in these galaxies (inside-out or outside-in, respectively). We show examples of all three morphological types in the sub-sections below. \par   

Once an initial identification of the edge is made using the above criteria, we plot an ellipse (the same used to derive the radial profiles) at {$R_{\rm edge}$} on the galaxy image to check whether the outskirts of the galaxy are elliptically symmetric in 2D. We show an example in the top right side of the flowchart of a galaxy and the profiles with the identified edge. If the galaxy is symmetric, we save all the relevant parameters in the edge, namely, $R_{\rm edge}$, the colour and mass density at the edge. \par

If a galaxy does not show a similar feature in its profile as shown in the example or is not symmetric in the outskirts, we further examine the 2D image for contaminants such as bright stars, neighbouring galaxies or streams that could affect the structure of the profile if the automated masking did not adequately remove the contaminated regions in the image. We also show an example in the lower section of the flowchart of a galaxy that contains tidal-like features beyond the edge. The profiles of this particular galaxy and other difficult cases are shown in Appendix \ref{app:difficult_cases}. We then re-compute the radial profiles for the galaxy and re-examine the data to locate an edge. We do not report edges in cases where even an improvement in masking and profile derivation did not allow us to locate an edge in our analysis (37\% of the galaxies analysed in this work; details in the next section.). \par 


\begin{figure*}[h!]
\centering
\begin{tikzpicture}[node distance=1.77cm]

\tikzstyle{startstop} = [rectangle, text width=3cm, minimum height=1cm,text centered, draw=black]
  \tikzstyle{io} = [trapezium, trapezium left angle=70, trapezium right 
  angle=110, text width=3cm, minimum height=1cm, text centered, draw=black]
  \tikzstyle{process} = [rectangle, text width=3cm, minimum height=1cm,    text centered, draw=black, dashed]
  \tikzstyle{decision} = [diamond, text width=3cm, minimum height=0.25cm, text centered, draw=black]
  \tikzstyle{arrow} = [thick,->, >=stealth]
  \tikzstyle{line} = [thick]

\tikzstyle{every node}=[font=\footnotesize]
\node (eachgalaxy) [process]{For each galaxy};
\node (inputs) [io, below of=eachgalaxy]{\textit{gri} composite image\\ $\mu_r$, $\mu_g$ \\ $g-r$  and \\ $\Sigma_{\star}(R)$ radial profiles};
\node (density_trunc) [decision, below of=inputs, aspect=2, yshift=-0.9cm]{Is there a signature of an edge in $\Sigma_{\star}(R)$?};

\node (rho_profile) [right of=density_trunc, xshift=8cm, yshift=0.1cm] 
{\includegraphics[width=.2\textwidth]{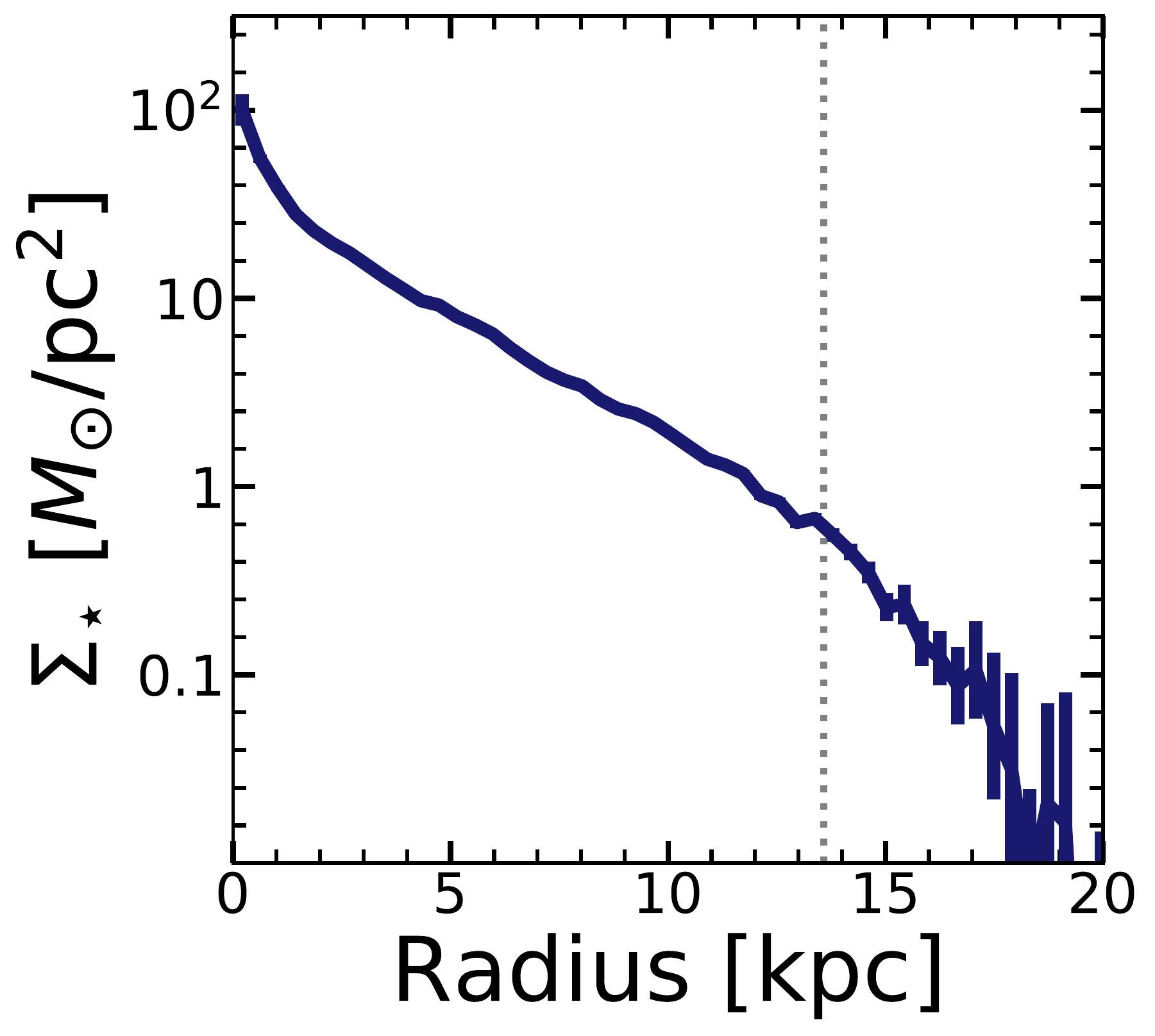}};

\node (sb_trunc) [decision, below of=density_trunc, aspect=2, yshift=-0.9cm]{Is there a signature of an edge in $\mu_g$ or $\mu_r$?};

\node (sb_profile) [right of=sb_trunc, xshift=8cm, yshift=0.1cm] 
{\includegraphics[width=.2\textwidth]{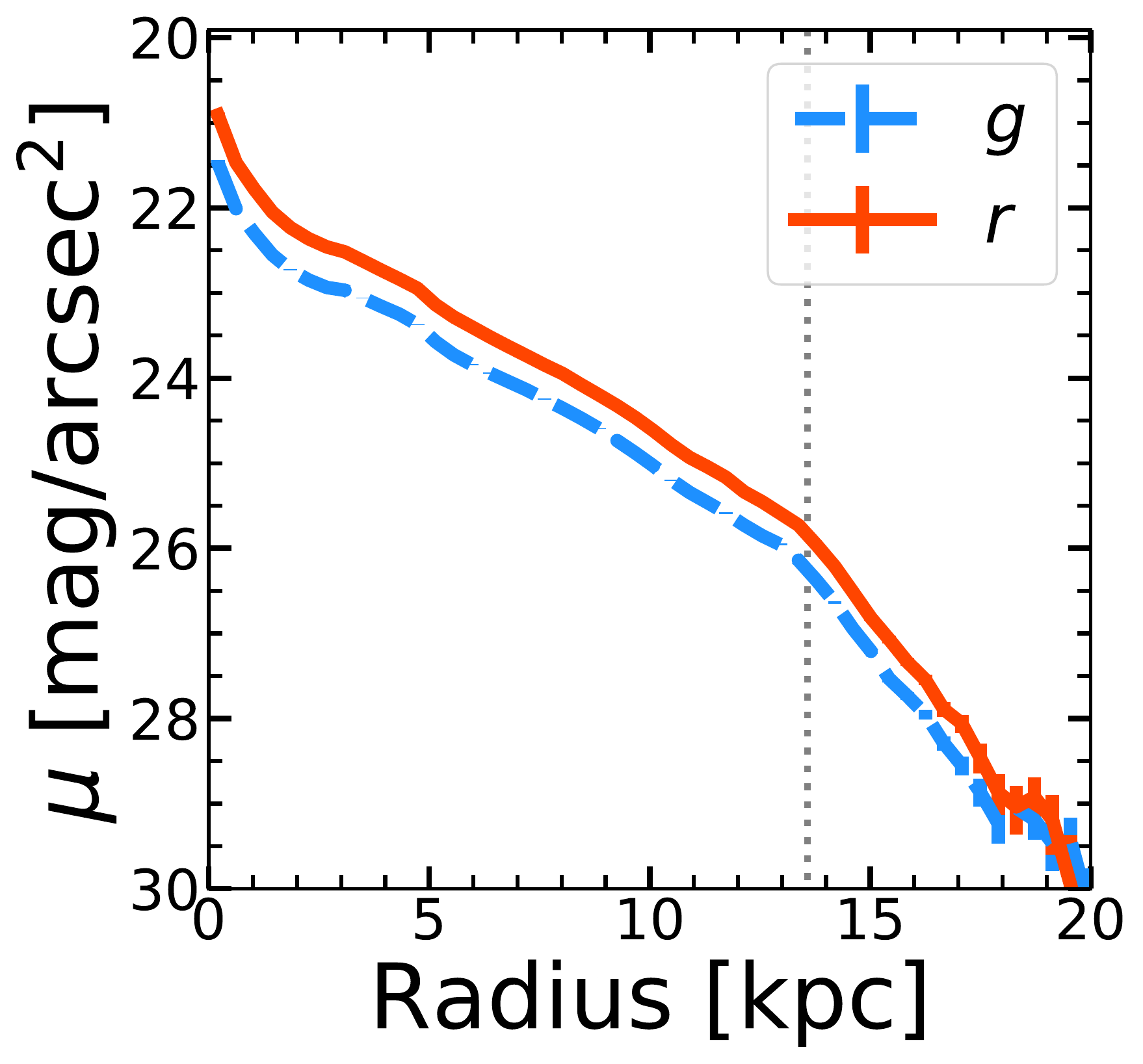}};

\node (sb_profile) [right of=sb_profile, xshift=2cm, yshift=0.15cm] 
{\includegraphics[width=.2\textwidth]{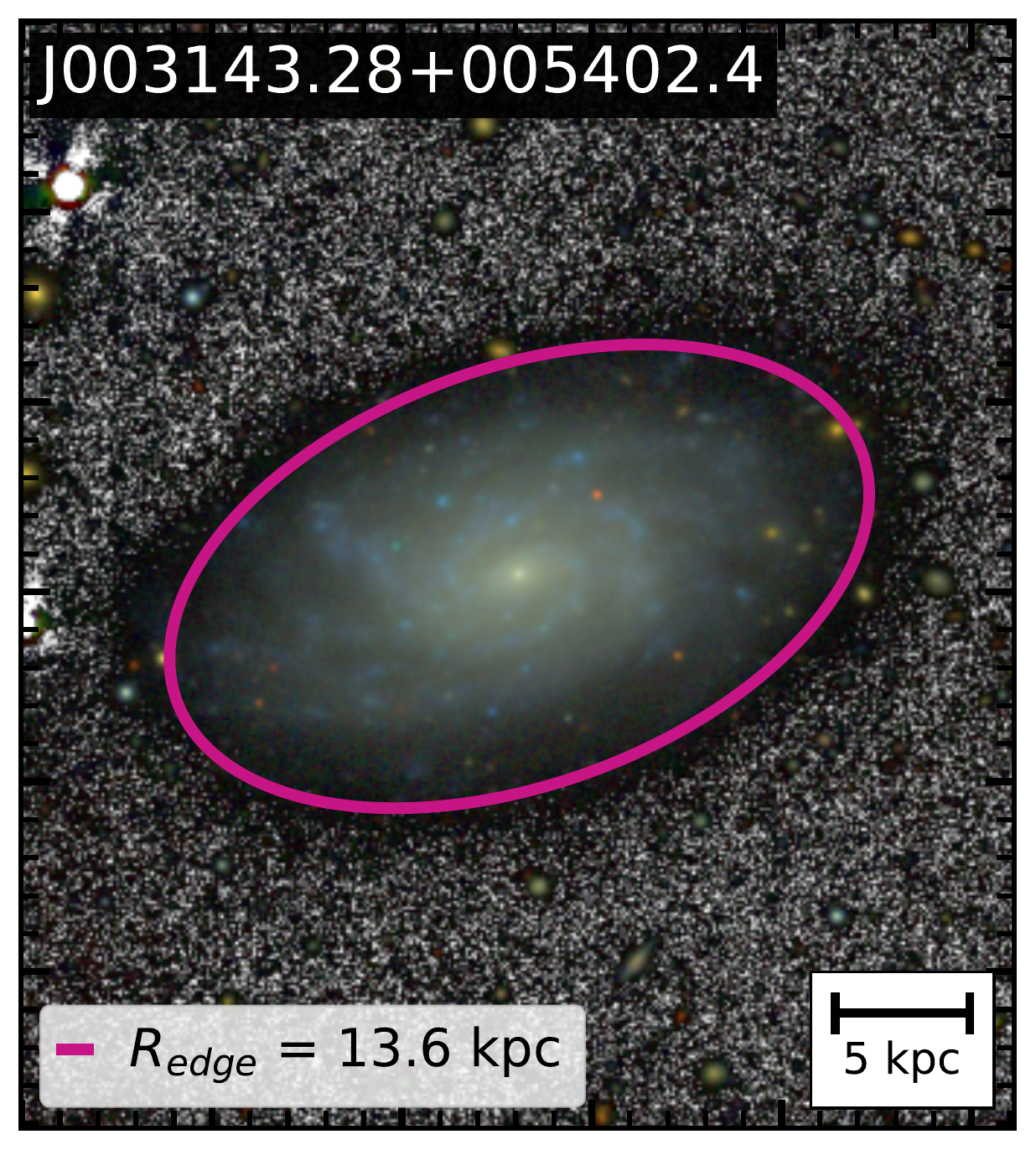}};

\node (color_trunc) [decision, below of=sb_trunc, aspect=2, yshift=-0.9cm]{Is there a signature of an edge in $g-r$?};

\node (col_profile) [right of=color_trunc, xshift=8cm, yshift=0.05cm] 
{\includegraphics[width=.2\textwidth]{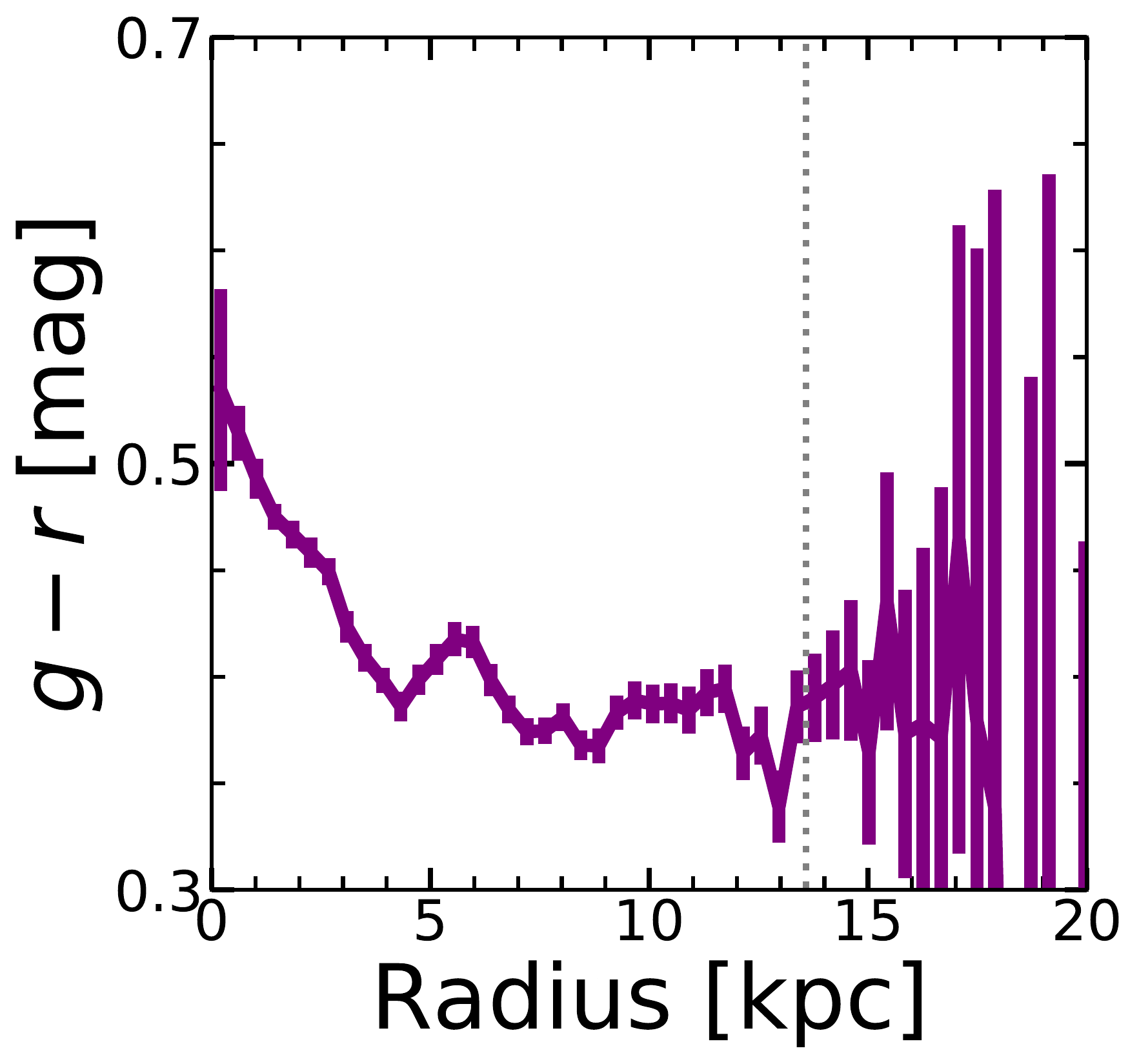}};

\node (plot) [startstop, right of=color_trunc, xshift=4 cm]{Plot ellipse at $R_{\rm edge}$ on galaxy image};
\node (within_galaxy) [decision, below of=plot, aspect=2, yshift=-0.9cm]{Are the outskirts (elliptically) symmetric (in 2D)?};
\node (save_params) [process, right of=within_galaxy, xshift=5cm, text width=4.5cm]{Save $R_{\rm edge}$, $(g-r)_{edge}$ and $\Sigma_{\star,edge}$};

\node (contamination) [startstop, below of=color_trunc, yshift=-3cm, text width=4cm]{Are there any non-symmetric features, including streams and contaminants (cirri, bright stars, neighbouring galaxies) in the image?};

\node (stream_fig1) [right of=contamination, xshift=6.5cm, yshift=-1cm] 
{\includegraphics[width=.2\textwidth]{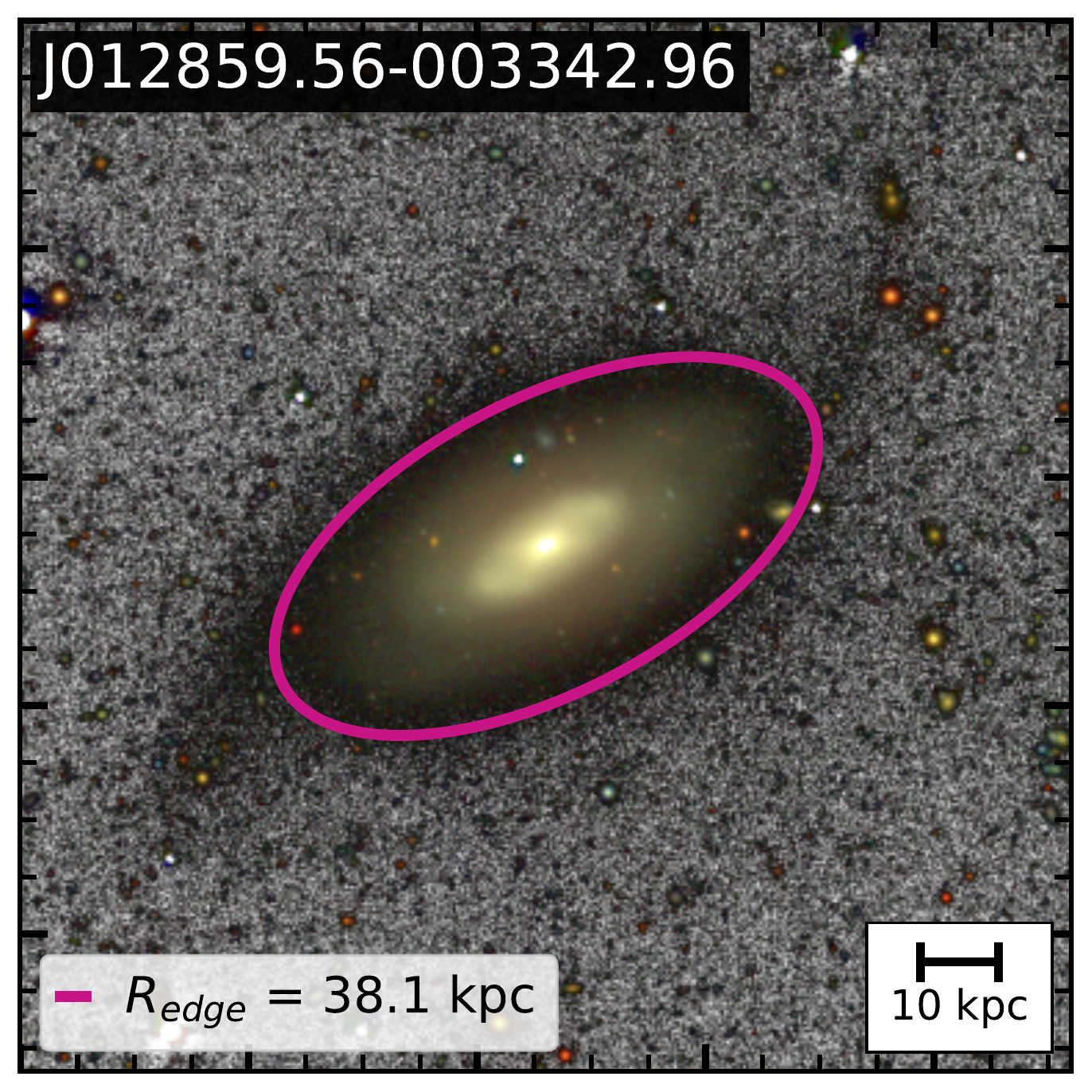}};

\node (improve_profiles) [decision, below of=contamination, yshift=-1.7cm]{Can the contaminants be removed or masked?};

\node (noedge) [startstop, below of=improve_profiles, yshift=-1.3cm]{No edge identified};

\draw [arrow] (eachgalaxy) -- (inputs);
\draw [arrow] (inputs) -- (density_trunc);
\draw [arrow] (density_trunc) -| node[anchor=south, xshift=-3cm] {Yes} (plot);
\draw [arrow] (sb_trunc) -| node[anchor=south, xshift=-3cm] {Yes} (plot);
\draw [arrow] (color_trunc) -- node[anchor=south] {Yes} (plot);
\draw [arrow] (plot) -- (within_galaxy);
\draw [arrow] (within_galaxy) -- node[anchor=south] {Yes} (save_params);
\draw [arrow] (within_galaxy) |- node[anchor=west] {No} (contamination);
\draw [arrow] (contamination) -- (improve_profiles);
\draw [arrow] (improve_profiles) -- node[anchor=west] {No} (noedge);
\draw [arrow] (contamination) -- node[anchor=west] {Yes} (improve_profiles);

\draw [arrow] (color_trunc) -- node[anchor=west] {No} (contamination);
\draw [arrow] (density_trunc) -- node[anchor=west] {No} (sb_trunc);
\draw [arrow] (sb_trunc) -- node[anchor=west] {No} (color_trunc);

\node (waypoint1) [minimum width=0, right of=improve_profiles]{};
\node (waypoint2) [right of=inputs, xshift=14cm]{};
\node (waypoint3) [left of=contamination]{};

\draw [line] (waypoint1.base) node[anchor=south, xshift=2cm] {Yes} -| (waypoint2.base);
\draw [arrow] (waypoint2.base) -- (inputs.east);
\draw [arrow] (waypoint3.base) |- node[anchor=south, xshift=-0.3cm] {No} (noedge);

\end{tikzpicture}
\caption{Flowchart illustrating the visual identification of edges in this work. An example of a galaxy, SDSS J003143.28+005402.4, and its profiles with the identified edge (vertical dotted line) is shown in the top right side of the flowchart.  If the galaxy is symmetric, the relevant parameters in the edge, namely, $R_{\rm edge}$, the colour and mass density at the edge are saved. However, if the galaxy contains any non-symmetric features such as tidal streams in its outskirts like the example SDSS J012859.56-003342.96 shown in the lower section of the flowchart, they are masked and the process of finding the edge is attempted once again. No edges are reported for cases when even an improvement in the masking did not present an edge. See Sect. \ref{sect:visual_procedure} for details.}
\label{fig:flowchart}
\end{figure*}
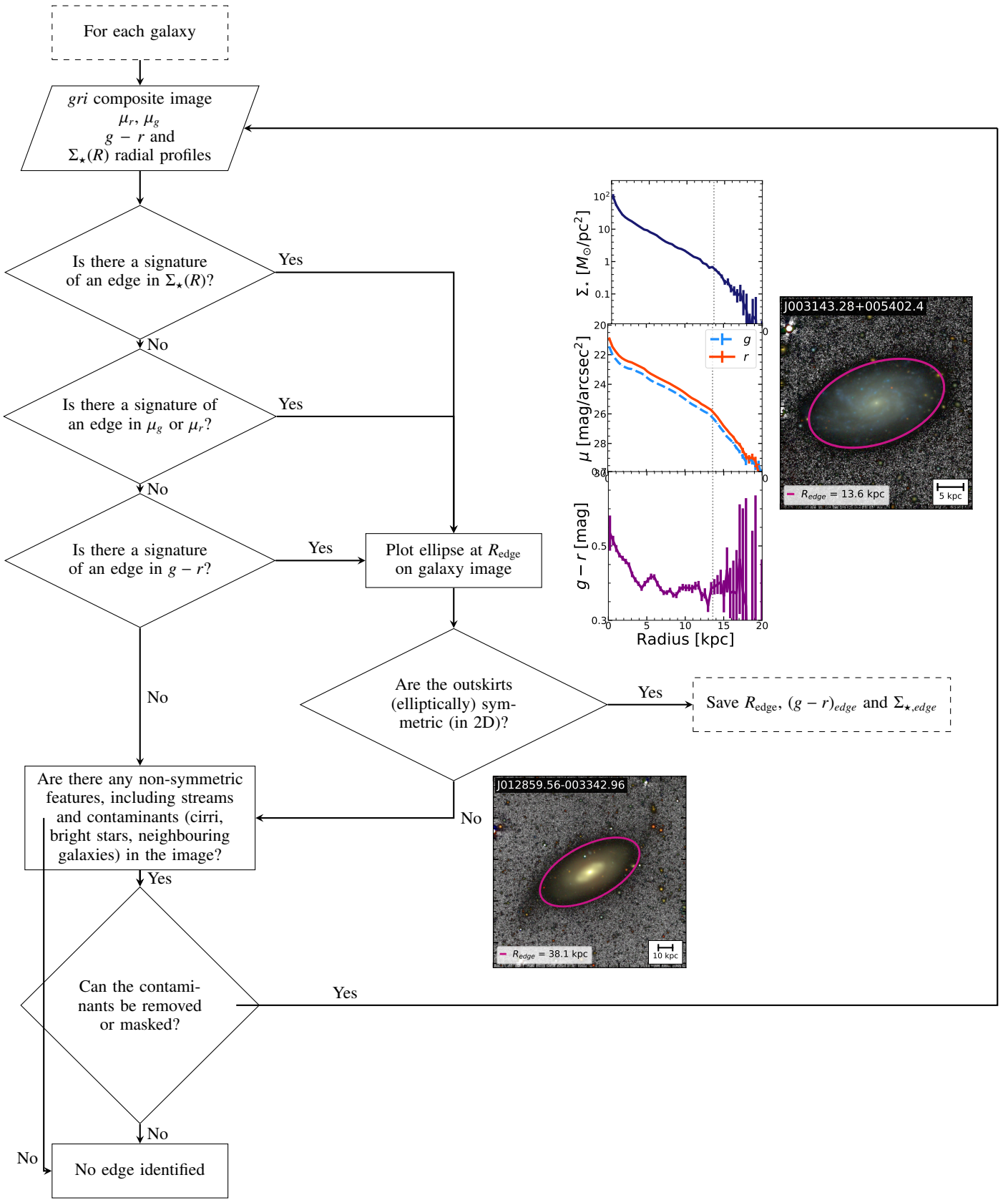

Following the procedure detailed above, the edge for each galaxy was identified by the authors N. Chamba and I. Trujillo. To quantify any dispersion in our inspections, we repeated our identifications and computed the average difference between these measurements. This quantity provides an estimate of the uncertainty in the visualisation of the edge. \par 

We also show that our criteria and measurements are independent of the depth of the imaging used here in Appendix \ref{app:image_depth}. The analysis is divided in two parts. In Appendix \ref{app:lbt}, we examine a nearby \citep[13.5 Mpc;][]{2019monelli} disk galaxy with an edge, NGC1042, using deeper imaging from the LIGHTS Survey \citep{2021trujillo}. We show that while deeper imaging allows one to characterise the edge with a higher signal-to-noise ratio, the edge of this galaxy may still be located with IAC Stripe 82 depth following our visualisation procedure. In Appendix \ref{app:limit_sb}, we compare the surface brightness at which the edges of our parent sample appear with the limiting surface brightness of the IAC Stripe 82 images used. We show that all the edges studied in this work appear at surface brightnesses above the limiting depth of our data. \par

\subsection{Dependence on orientation: late-type galaxies}
\label{sect:orient}

In the models studied by \citet{2014navarro}, the edge of a low-inclination Milky Way-like galaxy with a bulge, disk and stellar halo appears as a very soft bump (or shallow change in slope) in the outer surface brightness profile while that of an edge-on galaxy of equal stellar mass appears prominently as a sharper cut off. To complement this finding and visualise our definition of the edge, we show a few best case examples from our galaxy sample in Fig. \ref{fig:edges_example} and how the appearance of the edge of a galaxy changes with orientation for real late-type galaxies with similar (inclination corrected) rotational velocities (see Sect.  \ref{sect:sample}). For each galaxy, we show the IAC Stripe 82 $gri$-colour composite image with a pink contour marking the identified edge, $g$, $r$, GALEX NUV and FUV surface brightness profiles, $g-r$ colour profile and resulting stellar mass density profile, $\Sigma_{\star}$. The location of the edge is also marked in the panels as a dotted, grey vertical line. \par

Figure \ref{fig:edges_example} shows that edges appear at the location where there is a change in slope in the outer part of the surface brightness profiles (mainly in the UV). This location corresponds to the region where the $g-r$ colour rapidly becomes redder. The U-shape of colour profiles in disk galaxies were firstly identified in \citet{2008bakos} and \citet{2008azzollini}. This sudden reddening in the outer part of the colour profile is indicative of both the significant drop in in-situ star formation (as also confirmed by the truncation in the UV) and the emergence of the stellar halo component and/or stars migrated from the star forming regions in the disk to the outskirts. We make use of this feature collectively as the criteria to identify the edge in the rest of our late-type, spiral galaxy sample. \par 

In this work, the radial profiles of galaxies have been derived using elliptical annuli (Sect. \ref{sect:methods}).  In Appendix \ref{app:ellipse_semi}, we confirm that this method does not hinder our ability to identify the edges of low-inclination galaxies using their radial profiles compared to the method adopted for edge-on galaxies, i.e. using a slit through the galaxy's semi-major axis \citep[see e.g.][]{2019cristina}. We compare the ellipse and slit method for two galaxies: an edge-on case (UGC 09138) and a low-inclination one (NGC 1042) to study the effect of both these methods on galaxy orientation. We show that for UGC 09138, the elliptical annuli technique makes it unfeasible to locate the edge while for NGC 1042 the location of the edge is possible with either method. Therefore, our identification of the edge for low-inclination galaxies is not hampered by the method we adopt to derive their radial profiles in this work. \bigskip 

We illustrate and explain the criteria we use to identify the edges of the elliptical and dwarf galaxies in our sample in the sections below.

\subsection{The edges of early-type galaxies}
\label{sect:edges_ellipses}

Contrary to what happens in galaxies undergoing star formation, in the case of elliptical galaxies in our sample, their $g-r$ radial profiles follow a similar global shape as shown in Fig. \ref{fig:elliptical_examples}. The galaxies shown have a similar stellar mass $\sim 10^{11.5}\,M_{\odot}$. To locate their edges we have used a drastic change (a factor of five or higher difference in slope before and after the edge in the cases shown here) in colour in the outer parts of the system to mark a difference between the bulk of the object (with an homogeneous red colour) and potentially infalling (bluer) new material. Upon making this choice to mark the edge of these galaxies, we are implicitly assuming that the bulk of the elliptical galaxy was formed in an early burst and that the colour transition to the blue indicates the transition from the location of the original star formation radius to the outer envelope which was assembled more recently. \par

\subsection{The edges of dwarf galaxies}
\label{sect:edges_dwarfs}
For the dwarf galaxies in Fig. \ref{fig:dwarf_examples}, we find that the edge is visible in their $g-r$ profiles and/or stellar mass density profile. These galaxies have a stellar mass $\sim 10^8\,M_{\odot}$, and were chosen to illustrate the diversity in the colour radial profiles of dwarf galaxies which reflect their different morphology and substructure \citep[see also the work by][]{2016herrmann}. According to HyperLeda\footnote{\protect\url{http://leda.univ-lyon1.fr/}} \citep{2014makarov}, the morphology of these galaxies from case A to C labelled in the figure are Irr, Sd and SABd respectively.} The x-axis of the radial profiles are scaled to the location of $R_{\rm edge}$ and the located edge is marked with the vertical black line. The edges of these galaxies occur at mass densities $\Sigma_{\star}(R_{\rm edge}) \lesssim {2} M_{\odot}/$pc$^2$. For clarity, we additionally explore case A using the semi-major axis method in Appendix \ref{app:d23}, confirming the $R_{\rm edge}$ identified as a change in slope in the radial profiles. In case B and C particularly, the edge marks the transition towards either bluer or redder outskirts. This observation could be related to inside-out or outside-in star formation in the dwarfs \citep[e.g.][]{2012zhang}. In a future paper, we explore the connection between the transition to redder or bluer outskirts in dwarf galaxies and galaxy environment (Chamba \& Hayes in prep.). \par \bigskip 

\begin{figure*}[h!]
    \centering
    \includegraphics[width=1.0\textwidth]{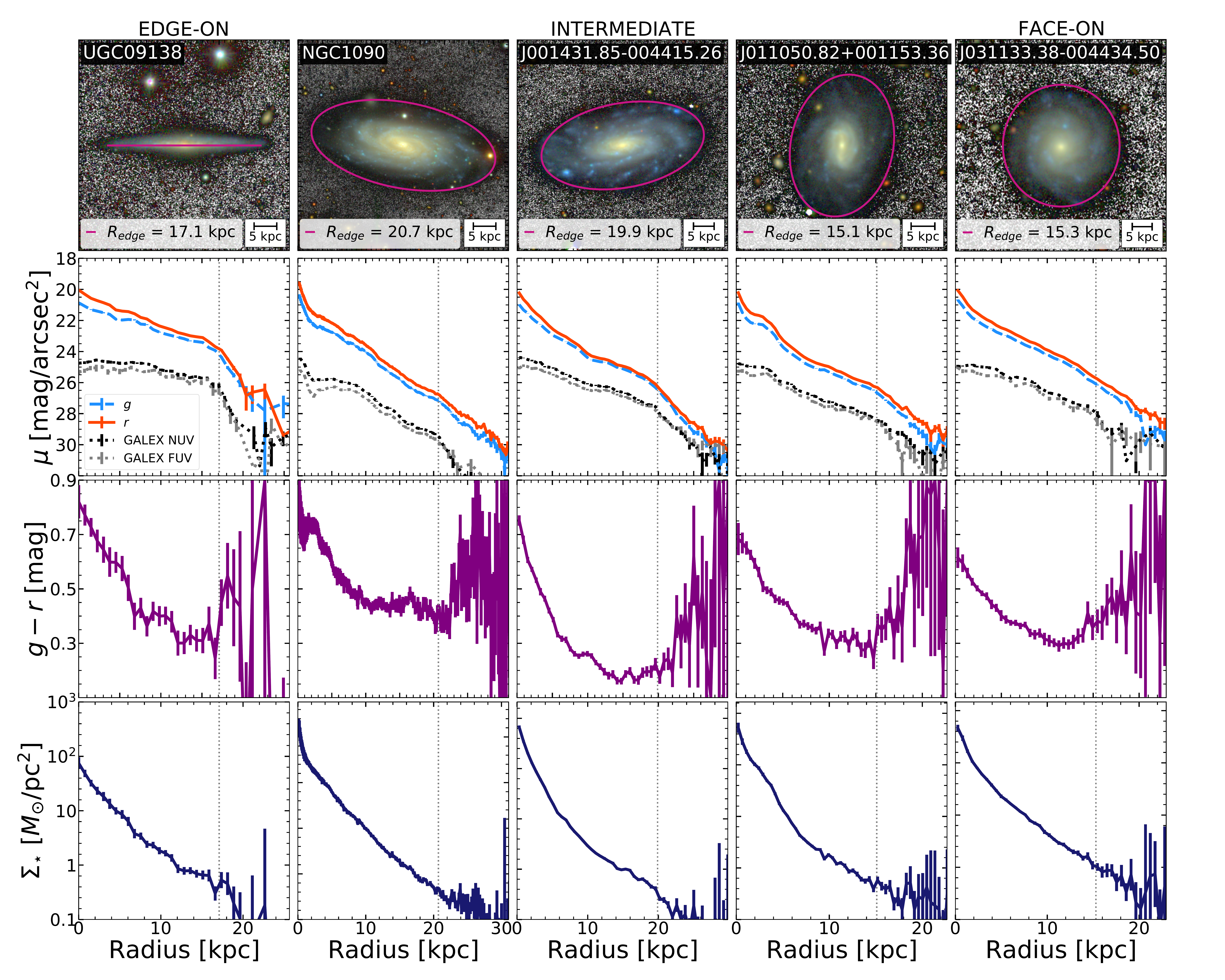}
    \caption{Edges of galaxies as viewed in edge-on (left) to face-on (right) orientations. The galaxies were selected to have similar rotational velocities $V_{rot} \sim 145\,$km/s. \textit{Left to right}: IAC Stripe 82 $gri$-colour composite image overlaid on a grey scaled $gri$-band summed image for contrast, surface brightness profiles in the SDSS $g$, $r$, GALEX NUV and FUV bands, $g-r$ colour profile and the corresponding $\Sigma_{\star}$ stellar mass density profile. The pink contour in the $gri$-image and the vertical dotted lines in the other panels indicate the edge of the galaxy.}
    \label{fig:edges_example}
\end{figure*}

\noindent We identify the edges of the spirals, ellipticals and dwarfs in the rest of our sample following the physically motivated criteria described above. In summary, the signature of the edge of a galaxy may be identified as:

\begin{itemize}
    \item[$\bullet$] a change in slope or cut-off in the radial surface brightness and/or stellar mass density profile (Sect. \ref{sect:concept}). 
    
    \item[$\bullet$] a sudden reddening in the outer part of the colour profile for spiral galaxies, indicative ofthe end of the star forming disk  (Sect. \ref{sect:orient}).
    
    \item[$\bullet$] a sudden transition from red to blue colours for elliptical  galaxies, to mark a difference between the core and recent infalling material, respectively (Sect. \ref{sect:edges_ellipses}).
    
    \item[$\bullet$] any of the above for dwarf galaxies. A transition to bluer or redder outskirts could reflect the inside-out or outside-in formation history, respectively (Sect. \ref{sect:edges_dwarfs}).
\end{itemize}

\noindent We leave the exploration of alternative criteria to mark the edges of galaxies for future work using deeper and/or higher resolution data.

\begin{figure*}
    \centering
    \includegraphics[width=0.8\textwidth]{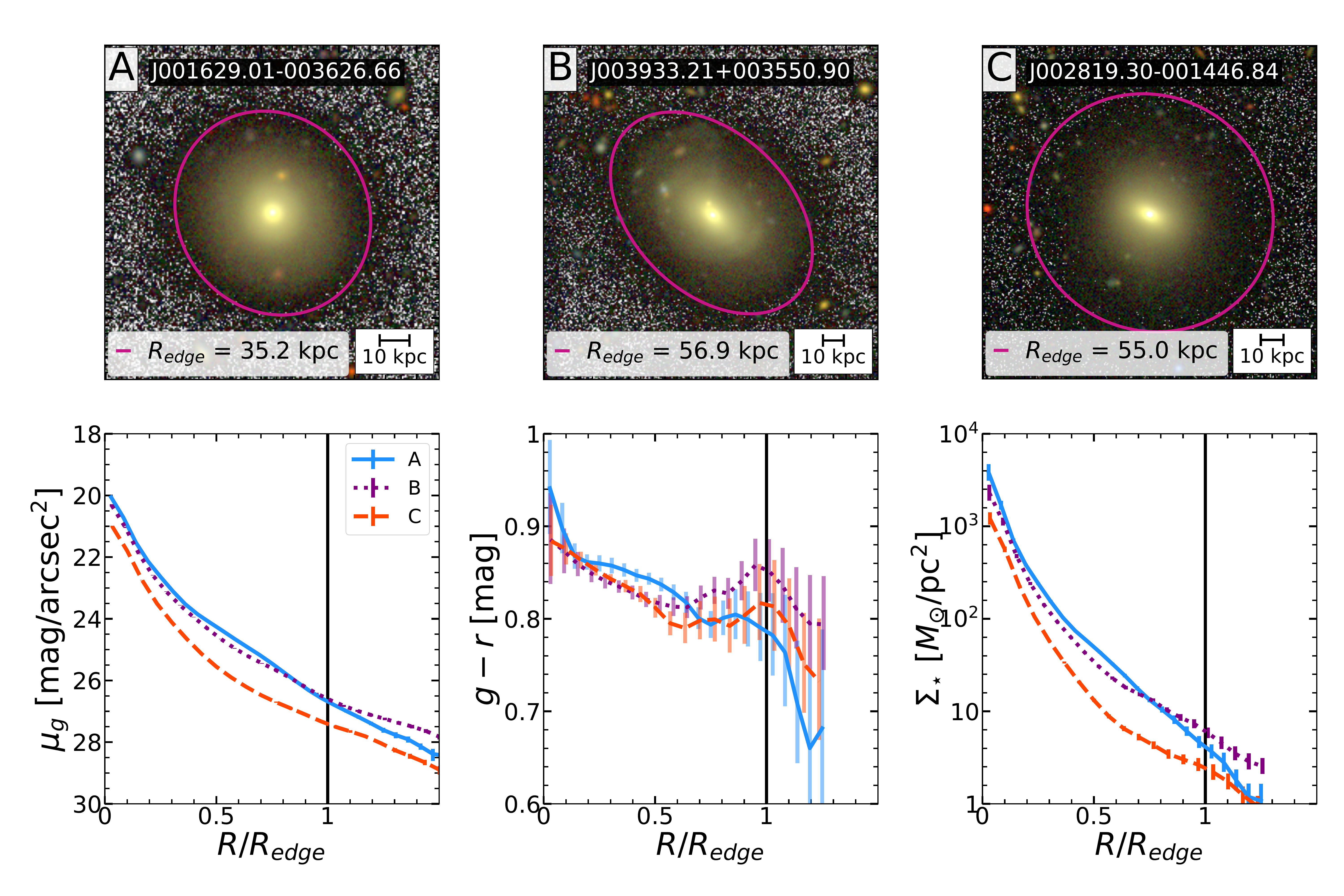}
    \caption{Three examples of elliptical galaxies from the parent sample with stellar mass $\sim 10^{11}\,M_{\odot}$. \textit{Top}: $gri$-band colour composite images, overlaid on the background $gri$ summed image in grey scale, of the galaxies denoted as A (left), B (middle) and C (right). The SDSS J2000 identifier and $R_{\rm edge}$ are labelled for these galaxies as in Fig. \ref{fig:edges_example}. \textit{Bottom, left to right}: The $\mu_g$, $g-r$ and $\Sigma_{\star}$ profiles of the objects. The edges for these galaxies are visible as a sudden transition from red to blue in their $g-r$ colour profiles (see Sect. \ref{sect:edges_ellipses} for details). The locations of the edges occur at mass densities $\Sigma_{\star}(R_{\rm edge}) > 1M_{\odot}/$pc$^2$.}
    \label{fig:elliptical_examples}
\end{figure*}

\begin{figure*}
    \centering
    \includegraphics[width=0.8\textwidth]{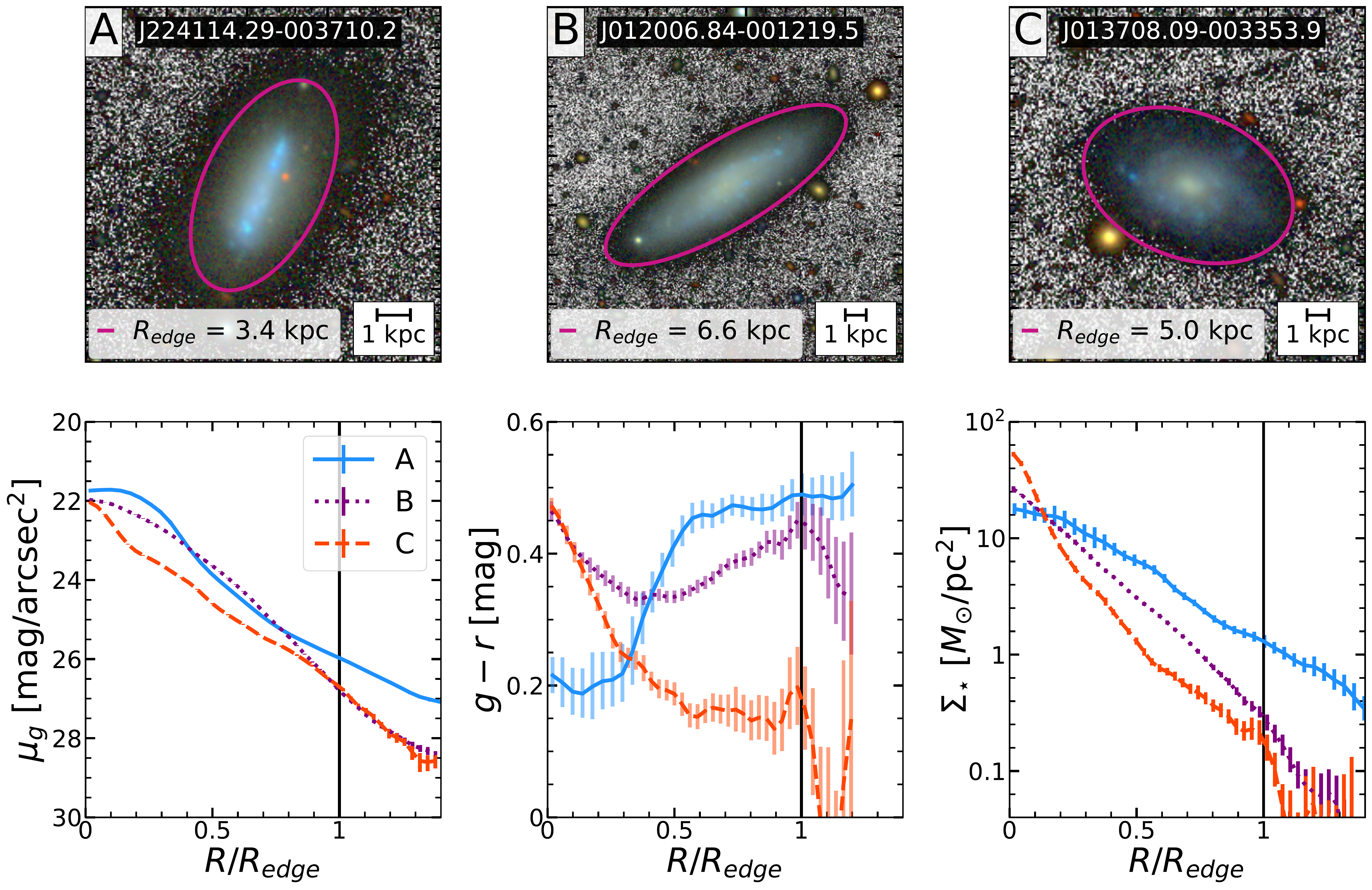}
    \caption{Similar to Fig. \ref{fig:elliptical_examples}, three dwarf galaxies with stellar mass $\sim 10^{8}M_{\star}$.These examples were specifically chosen to illustrate the diversity in the colour profiles of this galaxy population and the criteria we use to we locate the edge in these different cases. In case A, we use the change in slope in the $\Sigma_{\star}$ profile. In B and C, the edge is identified as a sudden transition to bluer colours in the $g-r$ profile (see Sect. \ref{sect:edges_dwarfs} for details). The edges of these galaxies occur at mass densities $\Sigma_{\star}(R_{\rm edge}) \lesssim {2}M_{\odot}/$pc$^2$.}
    \label{fig:dwarf_examples}
\end{figure*}

\clearpage 
\section{Results}
\label{sect:results}

\begin{figure*}
    \centering
    \includegraphics[width=0.95\textwidth]{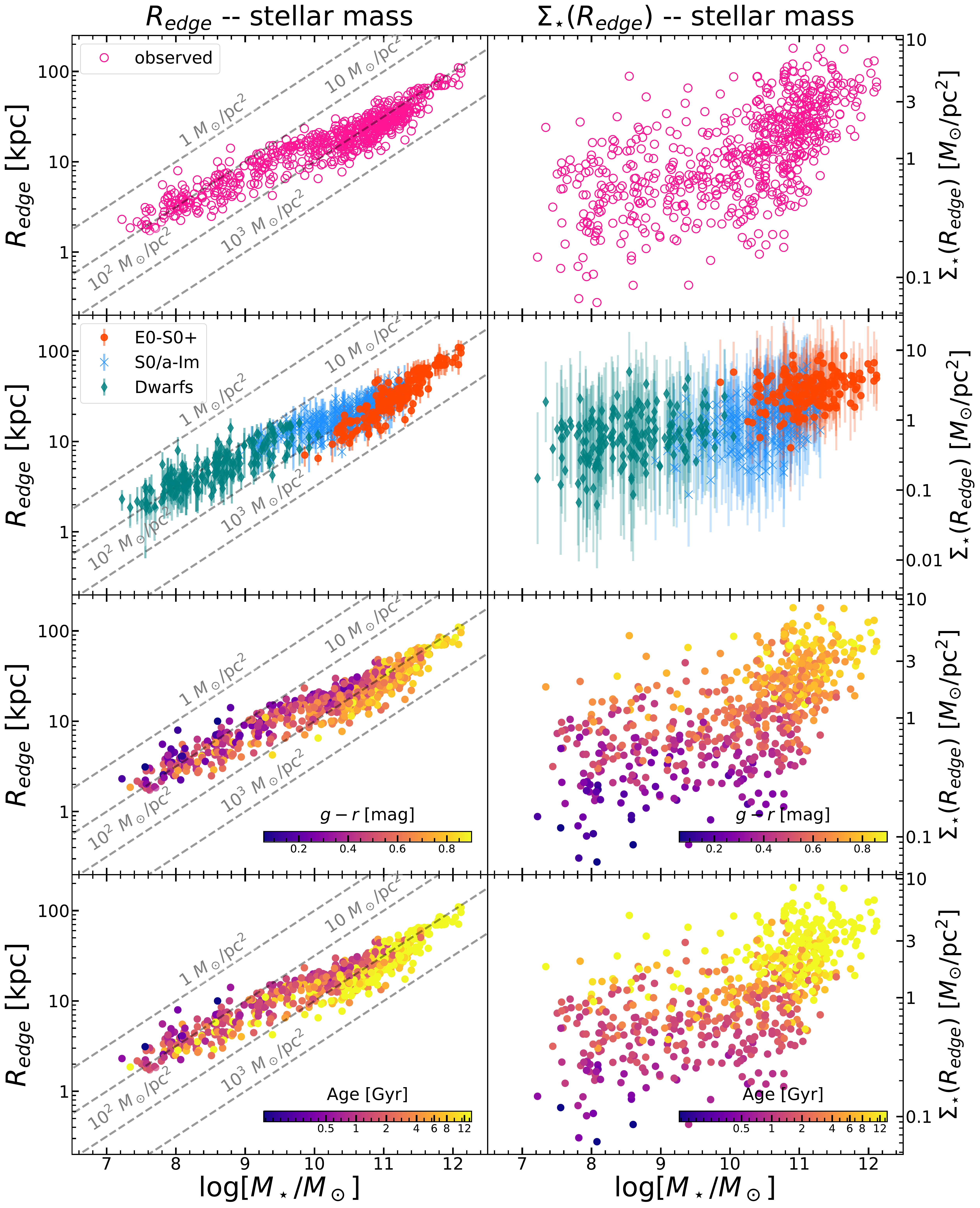}
    \caption{$R_{\rm edge}$--stellar mass (left) and $\Sigma_{\star}(R_{\rm edge})$--stellar mass (right) relations derived in this work. Only those galaxies where an edge was identified are plotted (624 objects). The grey lines in the $R_{\rm edge}$--stellar mass planes are lines of constant stellar mass surface density within the $R_{\rm edge}$ of the object. \textit{Top to bottom}: Each row shows the same observed relations (top), colour coded according to the morphology of the galaxies, ($g-r)_{edge}$ and a proxy for the age at $R_{\rm edge}$, for a fixed metallicity [M/H] = -0.71. We plot the uncertainties in our measurements (see Sect. \ref{sect:methods}) only in the second row for clarity in the other panels.}
    \label{fig:sigma-redge}
\end{figure*}

\begin{figure*}
    \centering
    \includegraphics[width=0.95\textwidth]{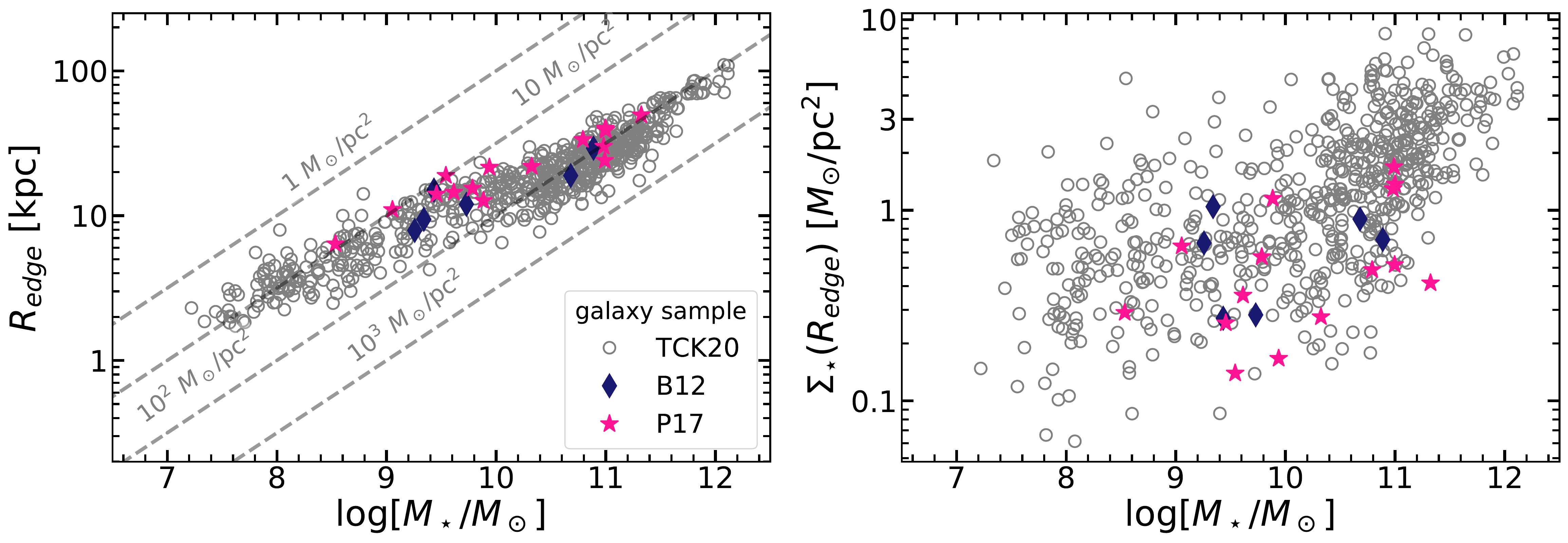}
    \caption{Similar to Fig. \ref{fig:sigma-redge}, now including the measurements in this work for the sample of galaxies studied in B12 and P17. The grey circles are the measurements for our sample.}
    \label{fig:sigma-redge-wnearby}
\end{figure*}

\begin{table*}[h!]
\centering
\caption{Best-fit parameters for the $R_{\rm edge}$-stellar mass relation. $\beta$ is the slope of the relation, $\sigma_{R_{obs}}$ is the dispersion, $\bar{\sigma}_{R_{obs}}$ is the observed dispersion corrected for visual identification errors (see text for details) and r is the Pearson correlation coefficient. }
\label{table:fitting} 
\begin{tabular}{c c c c c}
\hline 
Galaxy Type & $\beta$ & $\sigma_{R_{obs}}$ & $\bar{\sigma}_{R_{obs}}$ & r  \\  
 
\hline
\hline 

& & $R_{\rm edge}$--stellar mass & &  \\

\hline

All & 0.31$\pm$0.01 & 0.104$\pm$0.010 & 0.096 & 0.95  \\
E0-S0+ & 0.54$\pm$0.03 & 0.094 $\pm$0.013 & 0.092 & 0.91 \\
S0/a-Im & 0.27$\pm$0.02 & 0.097$\pm$0.058 & 0.093 & 0.82 \\
Dwarfs & 0.32$\pm$0.03 & 0.120 $\pm$0.012 & 0.118 & 0.85\\

\hline

\end{tabular}
\end{table*}

Using the above procedure, we identified edges in 171 (or 61.2\% of the) ellipticals, 273 (58.8\% of) spirals and 180 (68.7\% of) dwarfs (i.e 624 or 62\% of the galaxies in total) in our parent sample and 21 galaxies in the collective B12 and P17 samples. The 381 galaxies with no identified edges in our parent sample comprise of 107 ellipticals, 190 spirals and 84 dwarfs. 87 galaxies within this sub-sample were removed due to heavy contamination from bright stars, neighbouring galaxies and clouds of Galactic cirrus. Such highly contaminated galaxies may be studied with ad-hoc techniques but this is beyond the scope of the analysis and pipelines we developed for this work. Further investigation reveals that the majority of  galaxies without identified edges have very low inclination (mean axis-ratio of $q=0.71$), 84\% of which have $q > 0.5$. This finding is expected given the difficulty in identifying edges in low-inclination galaxies (see Fig. \ref{fig:edges_example}) . \par 

Figure \ref{fig:sigma-redge} shows the main result of this work: the $R_{\rm edge}$-stellar mass plane (left panels) and the $\Sigma_{\star}(R_{\rm edge})$-stellar mass plane (right panels) for the parent sample. Each row shows the data points in both planes labelled according to each galaxy's morphology, colour and age at the edge. The latter relations are similar when the total magnitudes of the galaxy in $g$ and $r$ are used to compute the colour and age. Our results also do not change if we compute the age with a fixed metallicity [$M/H$] of 0 or -0.71 (see Sect. \ref{sect:methods}). For clarity, we show our measurements for the nearby B12 and P17 galaxies separately in Fig. \ref{fig:sigma-redge-wnearby}. \par 

Following \citet{2020tck} and \citet{2020ctk}, we obtain the best fit slopes and dispersion values for the scaling relations using a Huber Regressor \citep{Huber1964} which is a linear regression model robust to outliers. The global $R_{\rm edge}$--stellar mass plane follows a power law of the form $R_{\rm edge} \propto M_{\star}^{\beta}$ where $\beta = 0.31\pm0.01$ and the relation has an observed dispersion of $\sigma_{R_{\rm edge}} = 0.10\pm0.01$ dex. For the individual galaxy populations, $\beta$ is $0.54\pm0.03$ for the elliptical galaxies (E0--S0+), $0.27\pm0.02$ for spirals (S0/a--Im) and $0.32\pm0.03$ for the dwarfs.  If we remove the uncertainty from our visual identification of $R_{\rm edge}$ ($\sigma_{\rm vis} \sim 0.04$\,dex) which we computed using NC and IT's repeated measurements (see Sect \ref{sect:visual_procedure}) from $\sigma_{R_{\rm edge}}$ in quadrature, we achieve a scatter of the relation $\bar{\sigma}_{R_{\rm edge}} = 0.096\pm 0.007$\,dex. These values are provided in Table \ref{table:fitting}. \par 

The above dispersion values also include observational errors due to background and stellar mass estimation. We find comparable values to the uncertainty in stellar mass in our galaxy sample to that published in \citet{2020tck} (i.e. 0.047 dex) and the effect of the background estimation on the location of $R_{\rm edge}$  amounts to 0.072 dex in our full sample. We may use these estimations of the uncertainty to compute the global intrinsic scatter of the size--stellar mass relation. Removing these values from $\sigma_{R_{\rm edge}}$ in quadrature  gives an intrinsic scatter of 0.059 dex. If we include the result of our visual identification of $R_{\rm edge}$ ($\sigma_{\rm vis} \sim 0.04$\,dex) the intrinsic scatter is 0.043 dex \citep[but see][for a detailed treatment of observational errors on the scatter of galaxy scaling relations]{2021stone}. \par

In the case of  the $\Sigma_{\star}(R_{\rm edge})$--stellar mass plane, we obtained two linear fits to describe the data by separating the sample in two intervals at $\log[{M_{\star}/M_{\odot}}]\sim 10.5$, i.e. interval $I_1$ where $\log[{M_{\star}/M_{\odot}}] < 10.5$ and $I_2$ where $\log[{M_{\star}/M_{\odot}}]\geq 10.5$. We split our sample in this fashion because 1) a single polynomial fit to the data performed very poorly and 2) spiral galaxies are over represented in our sample (274 objects out of the 624 galaxies with identified edges). Therefore, by splitting the sample at a stellar mass of $\log[M_{\star}/M_{\odot}]\sim 10.5$, the two intervals $I_1$ and $I_2$ are more comparable in terms of sample size (294 and 330 galaxies, respectively) and it is also the location where the slope of the $\Sigma_{\star}(R_{\rm edge})$--stellar mass relation increases. We plot these results explicitly in Appendix \ref{app:fits}. The two linear fits may be used to determine the average location of the edge (in mass density) $\left\langle \Sigma_{\rm edge}(M_{\star})\right\rangle$, as a function of galaxy stellar mass, given by:

\begin{align}
\label{eq:sigma-edge1}
    \log[\left\langle \Sigma_{\rm edge}(M_{\star})\right\rangle] =& 0.13\log \left [\frac{M_{\star}}{M_{\odot}} \right] - 1.32 \hspace{5mm} (\log \left [\frac{M_{\star}}{M_{\odot}} \right]< 10.5)\\
\label{eq:sigma-edge2}
    \log[\left\langle\Sigma_{\rm edge}(M_{\star})\right\rangle] =& 0.39\log \left[\frac{M_{\star}}{M_{\odot}} \right] - 3.97 \hspace{5mm} (\log \left[\frac{M_{\star}}{M_{\odot}} \right]\geq 10.5)
\end{align}

\noindent The uncertainties in the slopes  $\beta_{I_1} = 0.13\pm0.03$ and $\beta_{I_2} = 0.39\pm0.06$ and the dispersion in both relations are similar: $\sigma_{I_1} = 0.29\pm0.02$\,dex and $\sigma_{I_2} = 0.28\pm0.02$\,dex. \par

\begin{figure*}
    \centering
    \includegraphics[width=0.9\textwidth]{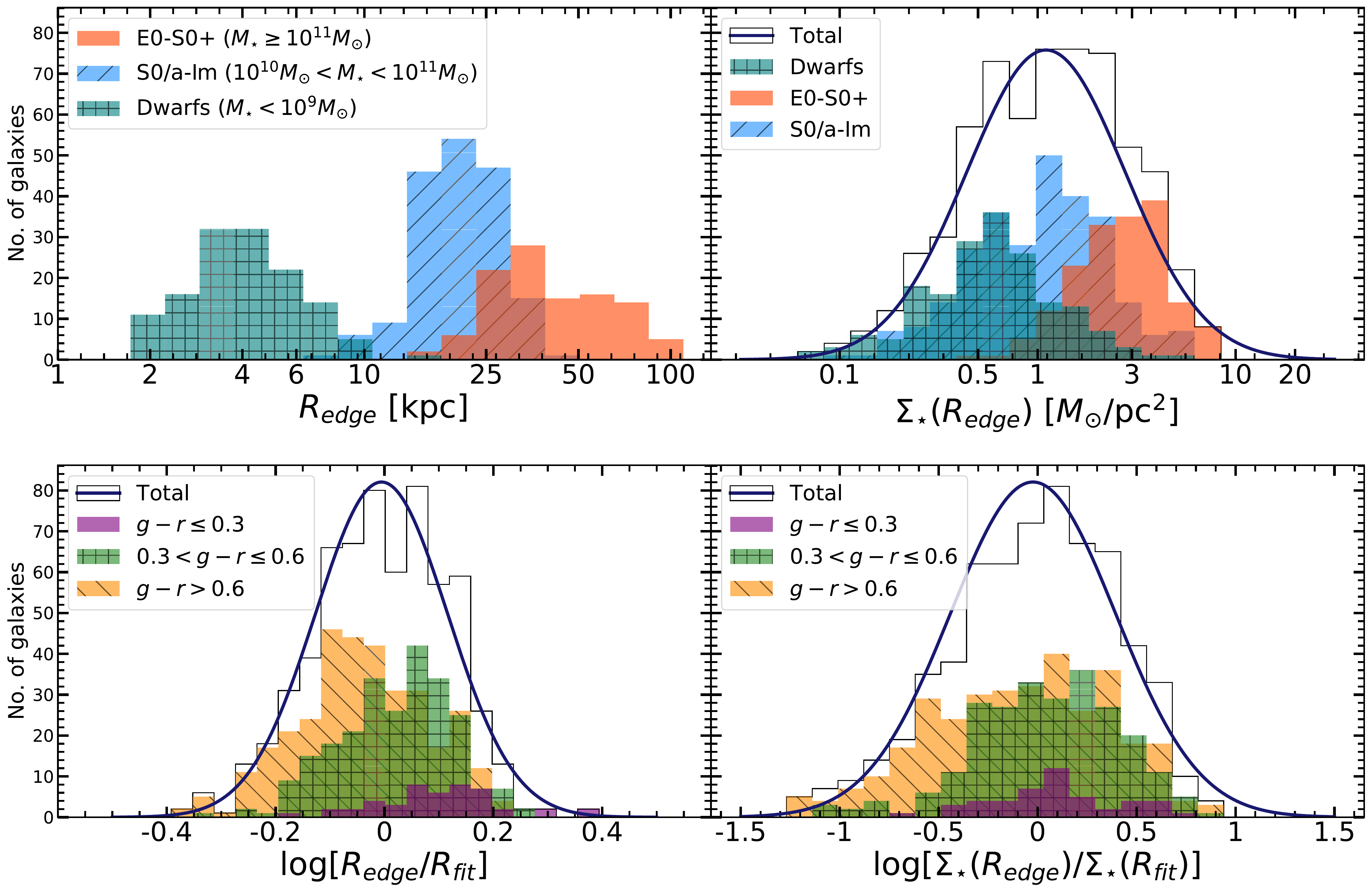}
    \caption{Representation of the results shown in Fig. \ref{fig:sigma-redge} as histograms. \textit{Top}: The distribution of $R_{\rm edge}$ (left) and $\Sigma_{\star}(R_{\rm edge})$. \textit{Bottom}: The distribution of ($g-r)_{edge}$ in the $R_{\rm edge}$–-stellar mass (left) and $\Sigma_{\star}(R_{\rm edge})$--stellar mass (right) relations. The subscript `fit' refers to the best fit line of each plane.}
    \label{fig:sigma-redge-hists}
\end{figure*}

We plot the distributions in $R_{\rm edge}$ and  $\Sigma_{\rm edge}$ as histograms in Fig. \ref{fig:sigma-redge-hists} for each morphological group studied here (upper panels) and we use the linear fits to the $R_{\rm edge}$-- and  $\Sigma_{\star}(R_{\rm edge})$--stellar mass planes to highlight the stratification in $(g-r)_{\rm edge}$ colour in those relations (lower panels). The subscript `fit' in this figure refers to the best fit line for each plane, i.e. the line that describes the average value of $R_{\rm edge}$ and $\Sigma_{\star}(R_{\rm edge})$ at each stellar mass. 

\begin{figure*}[t]
    \centering
    \includegraphics[width=1.0\textwidth]{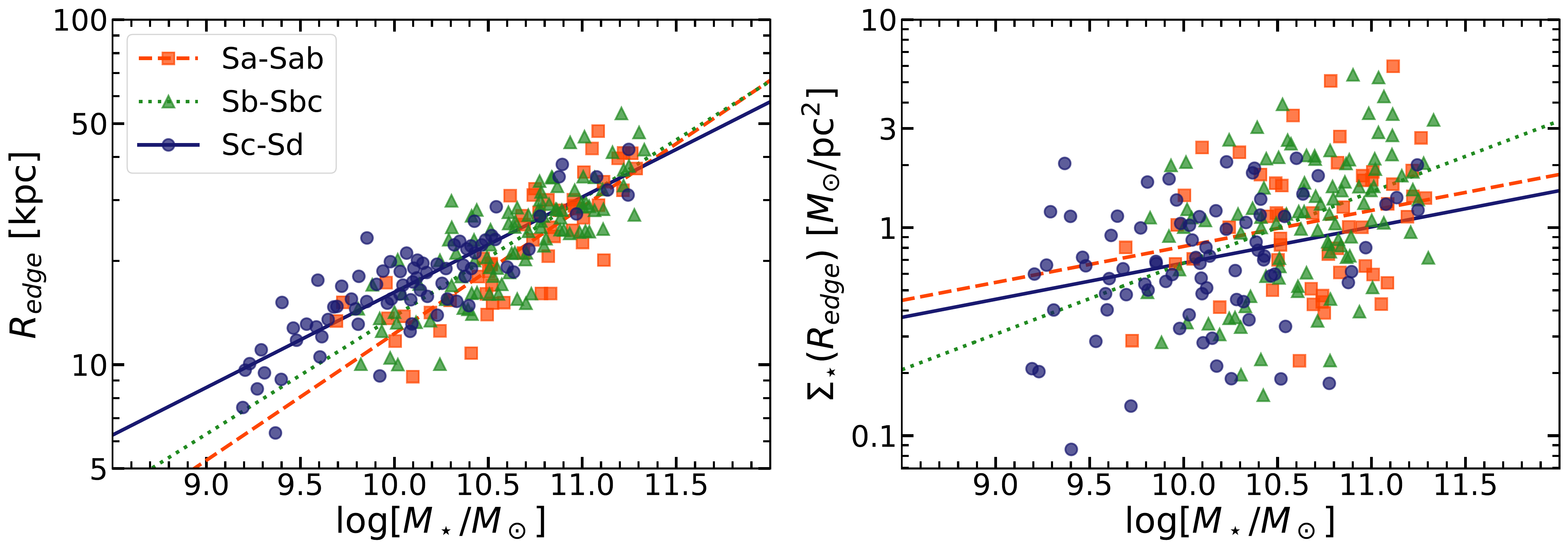}
    \caption{The stratification of late-type galaxies in the $R_{\rm edge}$--stellar mass (left) and $\Sigma_{\star}$($R_{\rm edge}$)--stellar mass (right) relation. The galaxies are grouped morphologically following \citet{2020tck}. The lines are the best-fit relations for each group in both panels. See text for details.}
    \label{fig:late_types}
\end{figure*}

The main features of the results shown in the above figures are as follows: 

\begin{itemize}

\item[$\bullet$] Galaxies where $M_{\star} \lesssim 10^{11}\,M_{\odot}$ closely follow a power law of the form $R_{\rm edge} \propto M_{\star}^{1/3}$ and is comparable to the global slope of the size--stellar mass relation. We additionally fit our data to a relation with a fixed
slope of 1/3, restricting the sample only to spiral galaxies or to both spiral and dwarf galaxies, finding that the y-intercept using both sub-samples did not change. This result supports the idea that both populations lie on the same slope. 

\item[$\bullet$] We observe a tilt in the $R_{\rm edge}$--stellar mass plane when $M_{\star} > 10^{11}\,M_{\odot}$. Massive elliptical galaxies dominate the scaling relation in this regime.  

\item[$\bullet$] The mass density at the location of the edge depends on the total stellar mass of the galaxy with an up turn at $\log[{M_{\star}/M_{\odot}}]\sim 10.5$. The slope of the $\Sigma_{\star}(R_{\rm edge})$--stellar mass plane triples at this stellar mass. 

\item[$\bullet$] In morphology, the average mass density at the location of the edge  is $2.9\pm0.1\,M_{\odot}/\text{pc}^2$ for the E0-S0+ sample, $1.1\pm0.04\,M_{\odot}/\text{pc}^2$ for S0/a-Im and $0.6\pm0.03\,M_{\odot}/\text{pc}^2$ for the dwarfs. 

\item[$\bullet$] The colour (age) gradient at fixed stellar mass with $R_{\rm edge}$ shows that larger galaxies have bluer (younger) edges.

\item[$\bullet$] The observed colour gradient also produces a gradient in mass density at the location of the edge at fixed stellar mass where bluer edges are located in regions of lower mass densities. More specifically, for galaxies within {a} stellar mass range of $10^{10}-10^{11}\,M_{\odot}$, we observe an age increase from $\sim$2 Gyr to $\sim$12 Gyr at the extreme ends of the scaling relation in $R_{\rm edge}$ as the size of the galaxy decreases, assuming a fixed metallicity of [M/H] = -0.71. 

\item[$\bullet$] On average, galaxies located in the upper half of both the $R_{\rm edge}$-- and $\Sigma_{\star}(R_{\rm edge})$--stellar mass relations have bluer edges compared to the lower half. In Appendix \ref{app:limit_sb} we show that this observation is not a bias due to image depth or the limiting surface brightness of our data. 

\item[$\bullet$] The B12 and P17 galaxies lie in the upper half of the $R_{\rm edge}$--stellar mass relation and in the lower regions of the $\Sigma_{\star} (R_{\rm edge})$--stellar mass relation. This is consistent with the fact that the majority of the galaxies in these samples have been classified with Sb, Sc or later morphology. {We plot the $R_{\rm edge}$-- and $\Sigma_{\star}(R_{\rm edge})$--stellar mass relations only for the late-type galaxies in our sample in Fig. \ref{fig:late_types} to highlight this statement}. 

\end{itemize}

\section{Discussion}
\label{sect:discuss}

We have visually identified the edges of a large sample of $\sim1000$ low-inclination galaxies spanning a wide morphology (from dwarfs to ellipticals) and stellar mass range ($10^7\,M_{\odot} < M_{\star} < 10^{12}\,M_{\odot}$). Sixty-two percent of the galaxies in our total sample presented identifiable edges following our visualisation procedure. We estimated the stellar mass density at their edge and then presented the resulting $R_{\rm edge}$-- and $\Sigma_{\star}(R_{\rm edge})$---stellar mass relations for these galaxies. \par 
Our main results are discussed in the following sub-sections. We leave the exploration of how our work may be used in future large-scale catalogues in Appendix \ref{sect:large_cats}. We find that the $\Sigma_{\star}(R_{\rm edge})$---stellar mass relations in Eqs. \ref{eq:sigma-edge1} and \ref{eq:sigma-edge2} could be used to obtain the location of the edge and provide a proxy for the size of any
galaxy, provided its stellar mass is known. These laws may be
useful for larger galaxy samples and automated cataloguing in
future multi-band surveys such as the LSST.   

\begin{figure*}[ht]
    \centering
    \includegraphics[width=0.95\textwidth]{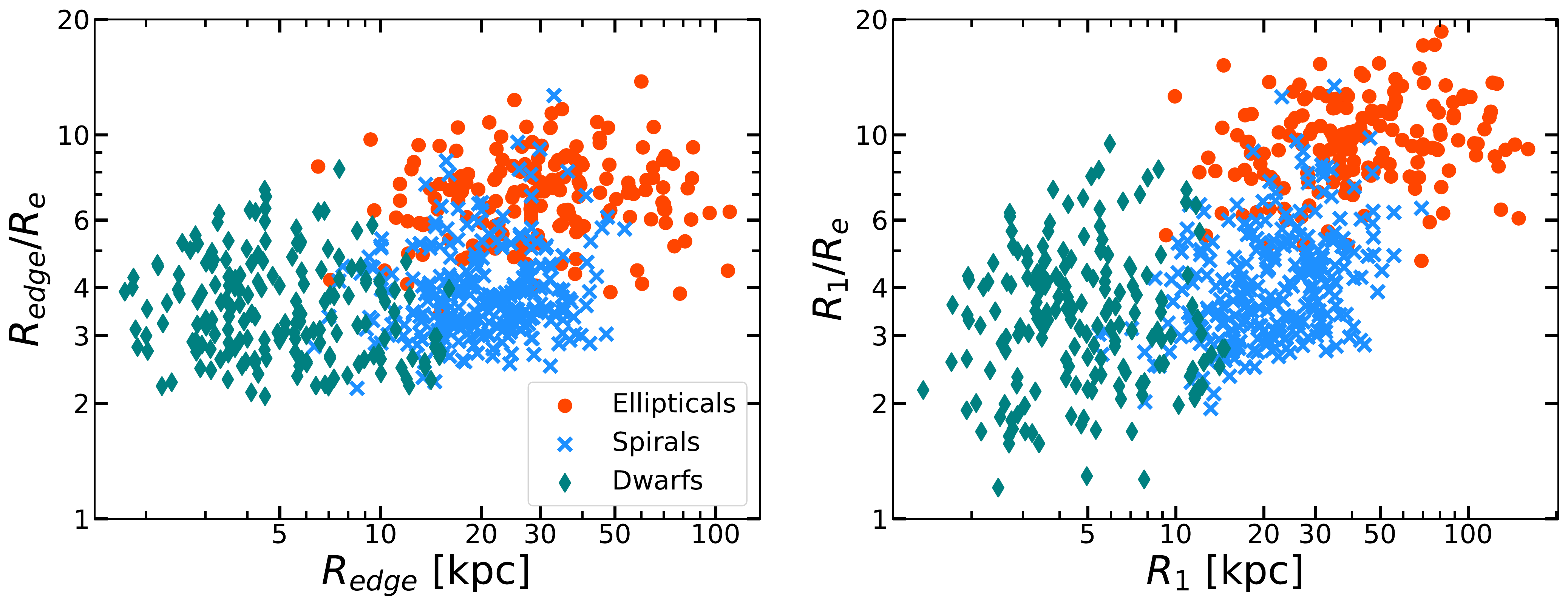}
    \caption{Ratios $R_{\rm edge}/R_{\rm e}$ (left) and  $R_{\rm 1}/R_{\rm e}$ (right) as a function of the respective sizes for the labelled morphologies. The diagrams reinforce how extended galaxies are compared to what the effective radius represents $R_{\rm e}$.}
    \label{fig:re_ratios}
\end{figure*}

\subsection{The global slope of size--stellar mass relation: $R_{\rm edge} \propto M_{\star}^{1/3}$}
\label{sect:global_slope}

If we focus on the spirals and dwarfs in our study, we have found that these galaxy populations have $R_{\rm edge}$--stellar mass relations with comparable slopes ($\beta \sim 0.3$ and close to a global slope $\sim 1/3$; see Table \ref{table:fitting}), with an intrinsic dispersion $\lesssim 0.06$\,dex. These parameters are compatible with those obtained in \citet{2020tck} using $R_1$ ({the fixed isomass contour at 1\,$M_{\odot}$/pc$^2$}) for size. The global slope of 1/3 is observed over four orders of magnitudes in stellar mass $10^7\,M_{\odot} < M_{\star} < 10^{11}\,M_{\odot}$ in the size--stellar mass relation, despite the different stellar mass surface densities we measured at the edges of galaxies in this regime. The lower density we measure at the edge for dwarfs ($\sim$0.6\,$M_{\odot}$/pc$^2$) compared to the spirals ($\sim$1\,$M_{\odot}$/pc$^2$) could be reflective of the low star formation efficiency that has been observed in these galaxies \citep[e.g.][]{2008leroy, 2012huang}. However, despite this difference, the constant slope could imply that the dwarfs and spiral galaxies in our sample share a common mechanism by which \textit{in-situ} star formation may have occurred. This idea is not incompatible with the fact that several similarities in the structure of surface brightness, colour and stellar mass density profiles of dwarfs and spiral galaxies have been found in the literature within the context of galaxy outskirts \citep[e.g.][]{2011hunter, 2016herrmann}. \par 

We contrast this result with those obtained using the effective radius \citep[$R_{\rm e}$;][]{1948dev}, a popular measure for galaxy size in the literature and with which the resulting size--stellar mass plane has very different characteristics. The relation is more broken for different galaxy morphologies \citep[see e.g.][]{2003shen, 2011brodie} and the dispersion is almost three times larger than that of the $R_{\rm edge}$--stellar mass plane \citep[see also][]{2020tck}. The broken scaling relations shown in these previous studies using $R_{\rm e}$ have traditionally been interpreted to reflect different size formation or evolution mechanisms for the galaxies. Therefore from this perspective, the global slope representing an unbroken size--stellar mass relation found here does not obviously favour such interpretations and works to readdress these issues are currently ongoing. However, to give the reader a view on how extended the galaxies in our sample are compared to what $R_{\rm e}$ suggests, we plot the ratios $R_{\rm edge}/R_{\rm e}$ and  $R_{\rm 1}/R_{\rm e}$ as a function of the respective sizes in Fig. \ref{fig:re_ratios}. {We compute $R_{\rm e}$ in a model independent way using the growth curve of the galaxy in the $g$-band which is our deepest dataset. The $R_1$ values were taken from \citet{2020tck}.} 

Particularly in $R_{\rm edge}$, it is clear for each galaxy type that while there is a change in $R_{\rm edge}$ over tens of kiloparsecs (e.g. 10-50\,kpc for spirals), the ratio with $R_{\rm e}$ is small (about 3--4). These results further support the need to rethink the concept of galaxy size and readdress the origin of the size--stellar mass relation \citep{2020sizeschamba}.

\subsection{Comparison between $R_{\rm edge}$ and the isomass contour at 1\,$M_{\odot}$/pc$^2$}
\label{sect:comp_r1}

\begin{figure}
     \centering
     \includegraphics[width=0.5\textwidth]{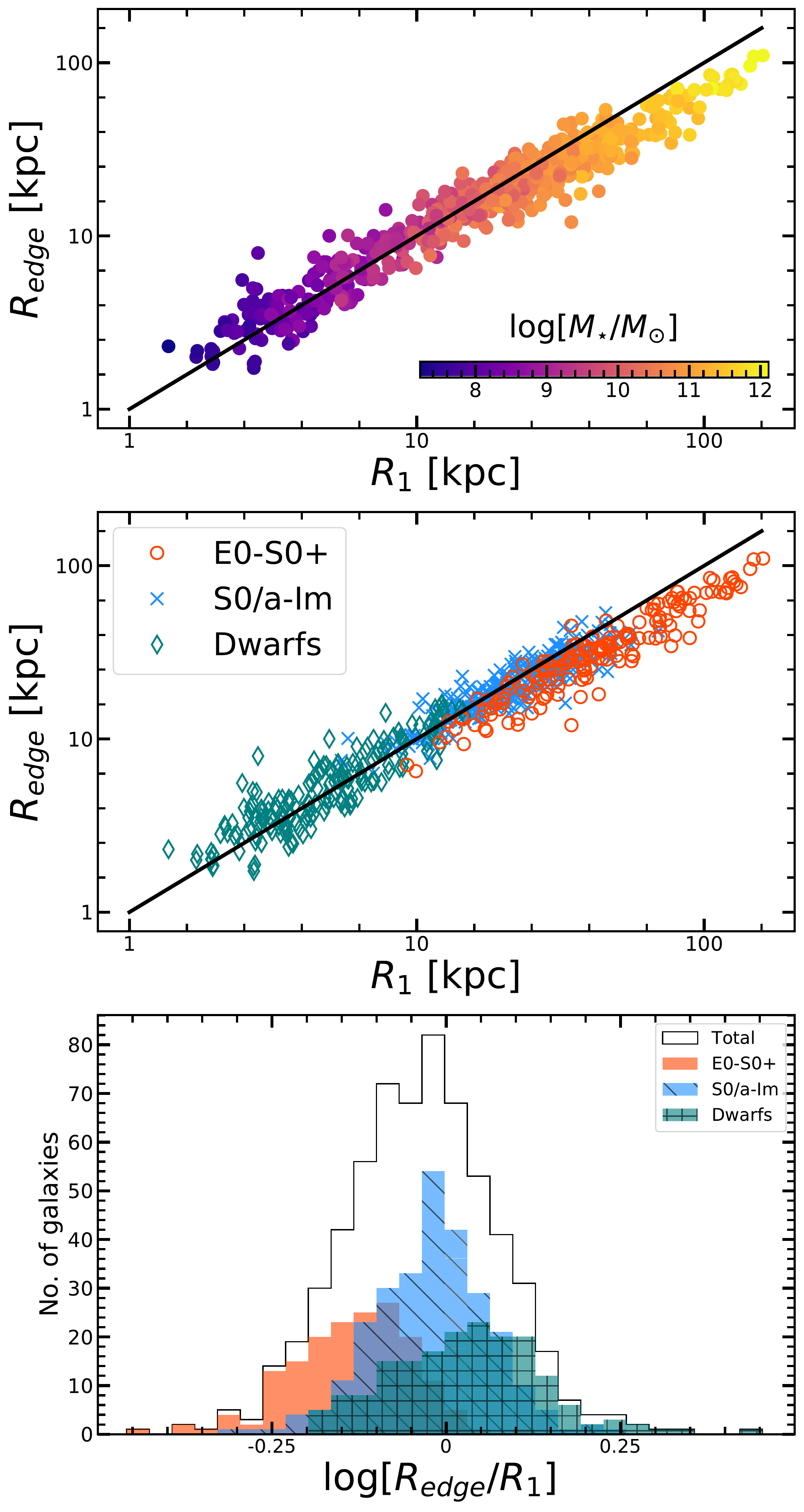}
     \caption{Comparison between $R_{\rm edge}$ and $R_1$. The top and middle panels show $R_{\rm edge}$ vs. $R_1$ for the galaxies in our sample, colour coded in stellar mass and morphology, respectively. The solid black line in the upper panels is the one-to-one relation. The lower panel shows the same distribution as a histogram.}
     \label{fig:comparison_w_r1}
 \end{figure}

To further address the structure we have observed in the size--stellar mass relation (stratification and slopes), it is worth exploring how similar $R_{\rm edge}$ is compared to $R_1$. As mentioned in the Introduction, we studied the $R_1$--stellar mass relation in \citet{2020tck} and \citet{2020ctk} because the isomass contour at $1 M_{\odot}/$pc$^2$ is physically motivated for Milky Way-like galaxies as it is related to a star formation threshold. A fixed isomass contour at $1 M_{\odot}/$pc$^2$ is also easier to measure\footnote{$R_1$ can be measured provided that the data is sufficiently deep, i.e. deeper than SDSS for low-inclination galaxies; see Fig. 6 in \citet{2021trujillo}.}, reproduce and is therefore more robust. For the sample of galaxies analysed here, we find that $R_1$ appears at a similar surface brightness to $R_{\rm edge}$:  on average $R_1$ appears at $\mu_g = 27.4\pm0.05$\,mag/arcsec$^2$ in $g$ (the standard deviation of this distribution is 1.2\,mag) and $R_{\rm edge}$ can be 0.4$\pm$0.02 \,mag/arcsec$^2$ brighter on average. Figure \ref{fig:comparison_w_r1} shows $R_{\rm edge}$ vs. $R_1$, colour coded according to the galaxy's stellar mass (top) and morphology (middle).  The lower panel shows the histogram of the distribution in the $R_{\rm edge}$--$R_1$ plane, with the sample divided according to their labelled morphology. The scatter in the $R_{\rm edge}$ vs. $R_1$ distribution is 0.088 dex with the elliptical (dwarf) galaxy population having smaller (slightly larger) $R_{\rm edge}$ than $R_1$ by $\sim 0.1$\,dex (0.05 dex) on average. Given these results, we recommend the use of $R_1$ for galaxies of similar properties as those studied in this work if the measurement of $R_{\rm edge}$ is not possible due to, for instance, poor signal-to-noise ratios. \par 

\begin{figure}
    \centering
    \includegraphics[width=0.5\textwidth]{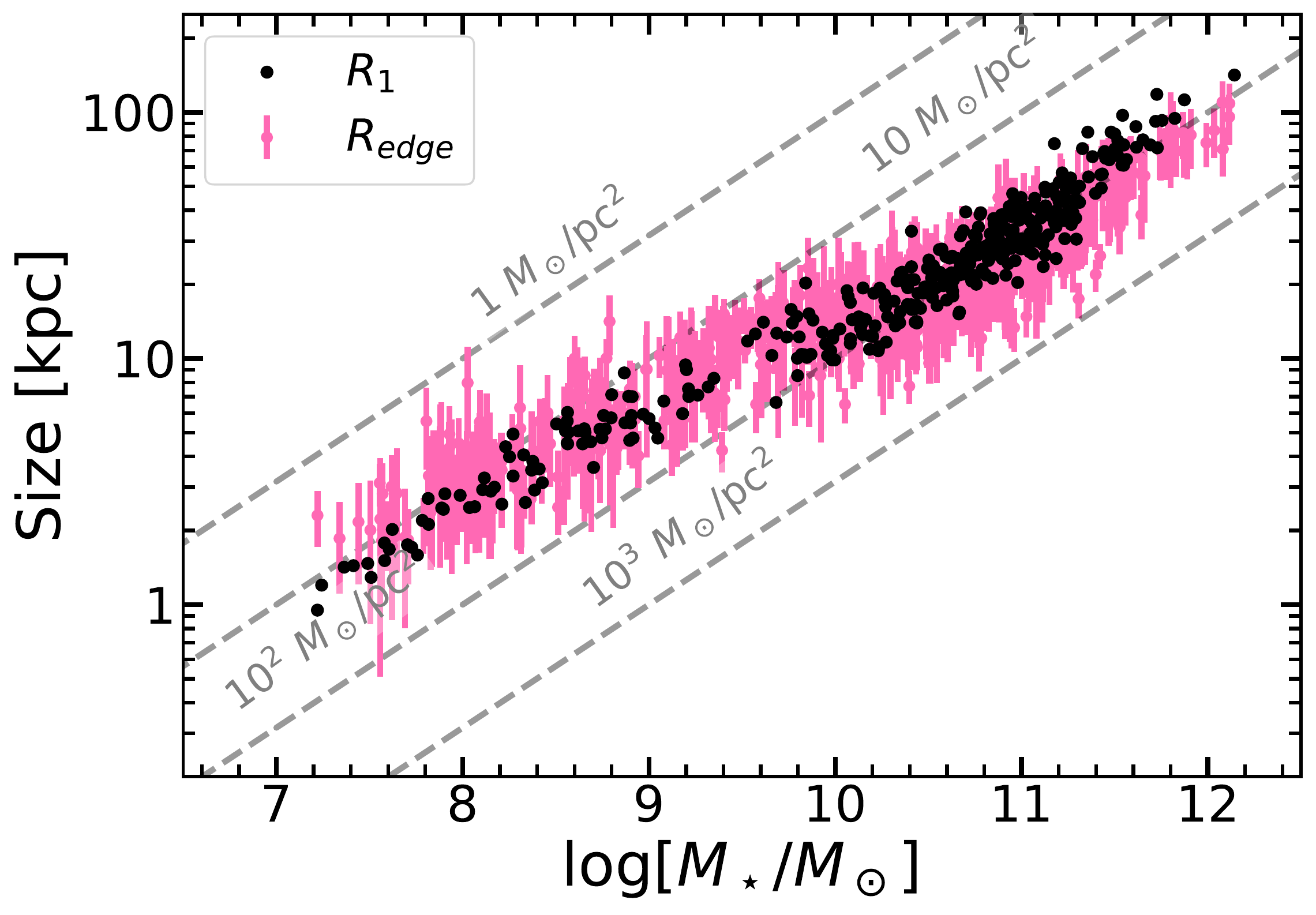}
    \caption{$R_1$ of galaxies where we {were unable to} locate the presence of edges (black) over plotted on the $R_{\rm edge}$--stellar mass plane (pink).}
    \label{fig:r1_removed_galaxies}
\end{figure}

Additionally, this result allows us to explore the $R_1$ of galaxies where we do not identify the presence of an edge. In Fig. \ref{fig:r1_removed_galaxies} we plot the $R_1$ of the 381 galaxies (37\% of the total sample) where we did not find the presence of edges in this work, the majority of which have very low inclinations or are heavily contaminated in their outskirts (see Sect. \ref{sect:results}) on the $R_{\rm edge}$--stellar mass plane. We see that the $R_1$ of these galaxies follow the overall distribution of $R_{\rm edge}$ in the size--mass plane within the uncertainties of our $R_{\rm edge}$ measurements. The most visible deviation occurs at the higher mass end ($M_{\star} > 10^{11}\,M_{\odot}$), however, this can be explained by the fact that the majority of our elliptical galaxies have $R_{\rm edge}$ at surface densities $> 1\,M_{\odot}$/pc$^2$ and thus results in smaller sizes (below the black points). In contrast, the $R_1$ of all of the dwarf galaxies shown here appear to be within the distribution of $R_{\rm edge}$, even though $R_{\rm edge}$ of the dwarfs are on average slightly larger (lower panel of Fig. \ref{fig:comparison_w_r1}). Therefore, from Fig. \ref{fig:r1_removed_galaxies} we may conclude that the exclusion of these galaxies without identified edges from our work does not bias our analysis and the main results discussed in this section  remain unchanged. However, we note that the scatter using $R_1$ is smaller compared to $R_{\rm edge}$ because of the difficulty in measuring the edge compared to a fixed isomass contour.

\subsection{The tilt in the $R_{\rm edge}$-- and $\Sigma_{\star}(R_{\rm edge})$--stellar mass planes}
\label{sect:tilt}

The comparison between $R_{\rm edge}$ and $R_1$ above shows that the two measures deviate from the one-to-one relation (solid black line in Fig. \ref{fig:comparison_w_r1}) for the majority of the higher mass elliptical galaxies in our sample (especially when $M_{\star} > 10^{11}\,M_{\odot}$), with $R_{\rm edge} < R_1$. Consequently, $R_{\rm edge}$ appears at stellar mass densities $> 1 M_{\odot}$/pc$^2$ ($\sim 3\,M_{\odot}$/pc$^2$ on average) for these galaxies and can be seen as a tilt in the $R_{\rm edge}$-- and $\Sigma_{\star}(R_{\rm edge})$--stellar mass relations (Fig. \ref{fig:sigma-redge}). {A similar tilt was observed in the $R_1$--stellar mass plane in \citet{2020tck}, however, here we have additionally found that the edges of these galaxies have older stellar populations than the late-types ($g-r$ colour > 0.6). The tilt observed here} (characterised by the slope of the size--stellar mass plane) is two times steeper for the ellipticals than the spirals and could be additional evidence towards a major difference in the mechanisms responsible for size growth between these two morphological groups, such as accretion which predominantly occurs in the most massive galaxies.\par 
To understand why the star formation threshold for elliptical galaxies are higher than for spiral galaxies, we need to consider the epoch at which these massive galaxies initially formed their stars. As pointed out in \citet{2020tck}, it has been observed that massive elliptical galaxies had bursty star formation histories at high redshift with star formation rates as extreme as 1000$\,M_{\odot}$/yr \citep[e.g.][]{2013riechers, 2015jaskot}. A high rate of star formation could provide a large enough energy budget to the surrounding gas, making it difficult for stars to form preferentially at low surface densities, i.e. increasing the galaxy's star formation threshold. Therefore, the edges of these massive galaxies could be tracing star formation that occurred during an epoch of high star formation rate which formed the core (bulge) material of the galaxy. Notice that $M_{\star} > 10^{11}\,M_{\odot}$ is also the regime where pressure supported systems could be dominating the scaling relation compared to rotationally supported galaxies \citep[see][]{2011emsellem}. This difference (as well as in symmetry, i.e. spherical vs. disc) between these galaxies could have consequences on the density at which stars would have preferentially formed in galaxies in the early Universe. \par 

In this work, we have selected a specific criterion to locate the edges of elliptical galaxies: a sudden colour transition from red to blue (Sect. \ref{sect:edges_ellipses}) and we have left the exploration of alternative criteria for future work. However, based on our results it is interesting to consider that deviations from the global slope in the size--stellar mass relation $R_{\rm edge} \sim M_{\star}^{1/3}$ could indicate deviations from the early star formation phase in the core component of these galaxies. If we fix the most massive galaxies with $M_{\star} > 10^{11.5}\,M_{\odot}$ (i.e. where only elliptical galaxies in our sample deviate from the global slope) to lie on the $R_{\rm edge} \sim M_{\star}^{1/3}$ relation, we obtain that their $R_{\rm edge}$ would be on average almost $\sim 20$\,kpc smaller. Consequently, the stellar mass density at these locations is more than double compared to that of the identified edges in this work for these galaxies (i.e $\sim 8\,M_{\odot}/pc^2$ on average). Such a threshold would enclose the bulk of the stellar component in relic galaxies, i.e. those galaxies which represent the core component of the majority of elliptical galaxies in the nearby Universe \citep[e.g. NGC 1277;][]{2014trujillo}. This alternative criterion would also better represent the visual edge of case C shown in Fig. \ref{fig:elliptical_examples}. With either this new criterion or that adopted in our identification procedure (Sect. \ref{sect:visual_procedure}), if the edges indicate a star formation threshold, our conclusion that the edges of elliptical galaxies occur at higher stellar mass densities remains unchanged.

\subsection{The stratification of late-types}
\label{sect:strat}
A stratification of late-type galaxies in morphology was previously found by \citet{2011spekkens} and \citet{2020tck} using $R_{23.5}$ (the isophote at 23.5 mag/arcsec$^2$ in the SDSS $i$-band) and $R_1$ respectively, with its origin still unknown. However, with $R_{\rm edge}$ we have found that while the average stellar mass density at the star formation threshold for these galaxies is $1\,M_{\odot}$/pc$^2$, there appears to be an additional variation in their stellar population properties where larger galaxies have bluer edges (Figs. \ref{fig:sigma-redge},  \ref{fig:sigma-redge-wnearby} and \ref{fig:late_types}). \par 

The stratification in size and colour suggests that larger galaxies may have reached their current size at a later time than smaller ones. This interpretation follows from the assumption that bluer colours trace ongoing or recent star formation while redder colours reflect the presence of older stellar material. For galaxies within the $10^{10}-10^{11}\,M_{\odot}$ stellar mass range studied here, outskirts that are populated with young to intermediate aged stars (< 6 Gyr) are possible due to, for example, the accretion of gas-rich satellite galaxies that trigger new star formation \citep[e.g.][]{2017grand}, the galaxy's enriched \textsc{Hi} content \citep[e.g.][]{2015kauffmann} or the migration of stars from the disk to the outskirts \citep[e.g.][]{2008roskar}. Of course the latter scenario could also result in redder outskirts with older stars but star formation may still occur in these regions \citep[see the review by][]{2017elmegreen}.  \par 

The idea that larger late-type galaxies are generally younger\footnote{The stratification in colour does not change whether we use the global $g-r$ of the galaxy or the colour at the edge.} is also not incompatible with the stratification we observe in their global morphology with Sa-Sab galaxies showing smaller sizes compared to Sc-Sd types (see Fig. \ref{fig:late_types}). Sa-Sab galaxies generally have lower star formation rates (often in a path towards quenching) than Sc or other later types where they are much higher \citep[e.g.][]{2016gonzalez, 2017bait}. Therefore, our data could be reflective of this  difference between galaxy types. Additionally, the lower half of the size--stellar mass relation is populated by elliptical galaxies of similar stellar mass (see Fig. \ref{fig:sigma-redge}). Elliptical galaxies are generally red, old and quiescent. From this perspective, the stratification we observe could be connected to when galaxies reached their present day stellar mass, i.e. a stratification between the `newer' late-type spirals and the `older' early-type elliptical galaxies \citep[see also the recent work by][]{2022watkins}. \par

The interpretation of whether a stratification in morphology exists in the lower mass dwarf regime is likely subjected to incompleteness effects \citep[{due to the limit in magnitude selected for spectroscopic targets in SDSS}; see the discussion in][]{2020ctk}. However, at least from the lower panels of Fig. \ref{fig:sigma-redge-hists}, there appears to be a stratification of galaxy size in colour where all of the blue edged dwarf galaxies ($g-r \lesssim 0.3$) appear in the upper half of the size--stellar mass relation. We do not possess morphological information of these galaxies in our sample to investigate this here further and leave it for future work. \par \bigskip

\section{Conclusions}
\label{sect:conclusions}

We have identified the edges of one of the largest samples of low-inclination galaxies. Our work expands on a physically motivated approach to define the edges (and consequently the sizes) of galaxies as the outermost location where {\textit{in-situ} star formation (either ongoing or in the past) significantly drops within these systems}. This idea is based on the gas density threshold required for the star formation process in galaxies \citep[e.g.][]{2004schaye}. Our main conclusions can be summarised as follows:

\begin{itemize}
\setlength\itemsep{0.5em}

     \item[$\bullet$] The size--stellar mass relation using $R_{\rm edge }$ has a global slope of $\sim 1/3$ and an intrinsic scatter $\lesssim 0.06$\,dex over a wide stellar mass range $10^7\,M_{\odot} < M_{\star} < 10^{12}\,M_{\odot}$ suggesting a common mechanism of \textit{in-situ} star formation. The structure of the relation is similar to that using $R_1$ \citep[see][]{2020tck, 2020ctk}.
     
     \item[$\bullet$] Massive elliptical galaxies dominate the scaling relation when $M_{\star} \gtrsim 10^{11}\,M_{\odot}$. This region corresponds to the tilt in the $R_{\rm edge}$-- and $\Sigma_{\star}(R_{\rm edge})$--stellar mass planes where the slope of the size--stellar mass relation doubles, potentially tracing the different epoch and high efficiency at which massive galaxies formed their stars.
     
     \item[$\bullet$] The stellar mass surface density at the edge (and consequently the star formation threshold) is a function of stellar mass and depends on the morphology of galaxies: it averages to $\sim 3\,M_{\odot}/$pc$^2$ ({or higher}) for ellipticals, $\sim 1\,M_{\odot}/$pc$^2$ for spirals and $\sim 0.6\,M_{\odot}/$pc$^2$ for dwarfs.

    \item[$\bullet$] $R_{\rm edge}$ is larger for bluer (i.e. younger) galaxies at a fixed stellar mass, reflective of when these galaxies reached their present-day size.
    
\end{itemize}

\noindent Given that $R_{\rm edge}$ is very similar to the location of the $1\,M_{\odot}/$pc$^2$ isomass contour ($R_1$) for the majority of the galaxies studied in this work, we recommend the use of the latter (or a higher isomass contour for elliptical galaxies) when the measurement of $R_{\rm edge}$ is challenging, for example due to imaging with low signal-to-noise. Due to the low scatter and physical significance underlying our measurements, we propose our size definition to be used in future deep, large-scale catalogues such as those from the LSST, to reach extreme galaxies of low surface brightness or high redshift galaxies with JWST and shift our understanding of how galaxies truly grow in size. 

\begin{acknowledgements}

We thank the anonymous referee for their detailed comments which helped to significantly improve the clarity of this manuscript. We acknowledge financial support from the State Research Agency (AEI-MCINN) of the Spanish Ministry of Science and Innovation under the grant `The structure and evolution of galaxies and their central regions' with reference PID2019-105602GB-I00/10.13039/501100011033 and grant PID2019-107427GB-C32, and from IAC projects  P/300624 and P/300724, financed by the Ministry of Science and Innovation, through the State Budget and by the Canary Islands Department of Economy, Knowledge and Employment, through the Regional Budget of the Autonomous Community. NC acknowledges support from the research project grant ‘Understanding the Dynamic Universe’ funded by the Knut and Alice Wallenberg Foundation under Dnr KAW 2018.0067 and thanks Matthew Hayes, Fernando Buitrago, Claudio Dalla Vecchia, Aura Obreja, L. Pascho, and Naomi Samsoodeen for interesting discussions. J.H.K. and I.T.C. acknowledge support from the ACIISI, Consejer\'{i}a de Econom\'{i}a, Conocimiento y Empleo del Gobierno de Canarias and the European Regional Development Fund (ERDF) under grant with reference PROID2021010044.\\ 

Funding for the Sloan Digital Sky Survey IV has been provided by the Alfred P. Sloan Foundation, the U.S. Department of Energy Office of Science, and the Participating Institutions. SDSS-IV acknowledges
support and resources from the Center for High-Performance Computing at the University of Utah. The SDSS web site is www.sdss.org. SDSS-IV is managed by the Astrophysical Research Consortium for the Participating Institutions of the SDSS Collaboration including the Brazilian Participation Group, the Carnegie Institution for Science, Carnegie Mellon University, the Chilean Participation Group, the French Participation Group, Harvard-Smithsonian Center for Astrophysics, Instituto de Astrof\'isica de Canarias, The Johns Hopkins University, Kavli Institute for the Physics and Mathematics of the Universe (IPMU) / University of Tokyo, the Korean Participation Group, Lawrence Berkeley National Laboratory, Leibniz Institut f\"ur Astrophysik Potsdam (AIP), Max-Planck-Institut f\"ur Astronomie (MPIA Heidelberg), Max-Planck-Institut f\"ur Astrophysik (MPA Garching), Max-Planck-Institut f\"ur Extraterrestrische Physik (MPE), National Astronomical Observatories of China, New Mexico State University, New York University, University of Notre Dame, Observat\'ario Nacional / MCTI, The Ohio State University, Pennsylvania State University, Shanghai Astronomical Observatory, United Kingdom Participation Group, Universidad Nacional Aut\'onoma de M\'exico, University of Arizona, University of Colorado Boulder, University of Oxford, University of Portsmouth, University of Utah, University of Virginia, University of Washington, University of Wisconsin, Vanderbilt University, and Yale University. \\
This research has made use of the NASA/IPAC Extragalactic Database (NED), which is operated by the Jet Propulsion Laboratory, California Institute of Technology, under contract with the National Aeronautics and Space Administration. \\
We acknowledge the usage of the HyperLeda database \url{http://leda.univ-lyon1.fr} \\
This work was partly done using GNU Astronomy Utilities (Gnuastro, ascl.net/1801.009) version 0.14. Work on Gnuastro has been funded by the Japanese Ministry of Education, Culture, Sports, Science, and Technology (MEXT) scholarship and its Grant-in-Aid for Scientific Research (21244012, 24253003), the European Research Council (ERC) advanced grant 339659-MUSICOS, European Union’s Horizon 2020 research and innovation programme under Marie Sklodowska-Curie grant agreement No 721463 to the SUNDIAL ITN, and from the Spanish Ministry of Economy and Competitiveness (MINECO) under grant number AYA2016-76219-P. \par \newline

\textit{Software}: \texttt{Astropy},\footnote{http://www.astropy.org} a community-developed core \texttt{Python} package for Astronomy \citep{robitaille2013astropy, price2018astropy}; \texttt{Gnuastro} \citep[astwarp;][]{2015gnuastro};  \texttt{IMFIT} \citep{2015ApJ...799..226E}; \texttt{Jupyter Notebooks}  \citep{Kluyver:2016aa}; \texttt{Matplotlib} \citep{Hunter:2007}; \texttt{MTObjects} \citep{2016mto, 2021haigh}; \texttt{NumPy} \citep{numpy, doi:10.1109/MCSE.2011.37}; \texttt{SAO Image DS9} \citep{ds9}; \texttt{SciPy} \citep{scipy}; \texttt{SWarp} \citep{swarp}; and \texttt{TOPCAT} \citep{2005ASPC..347...29T}.

\end{acknowledgements}

\bibliographystyle{aa}
\bibliography{bibliography.bib}

\appendix

\section{Dealing with difficult cases}
\label{app:difficult_cases}

\begin{figure*}
    \centering
    \includegraphics[width=0.9\textwidth]{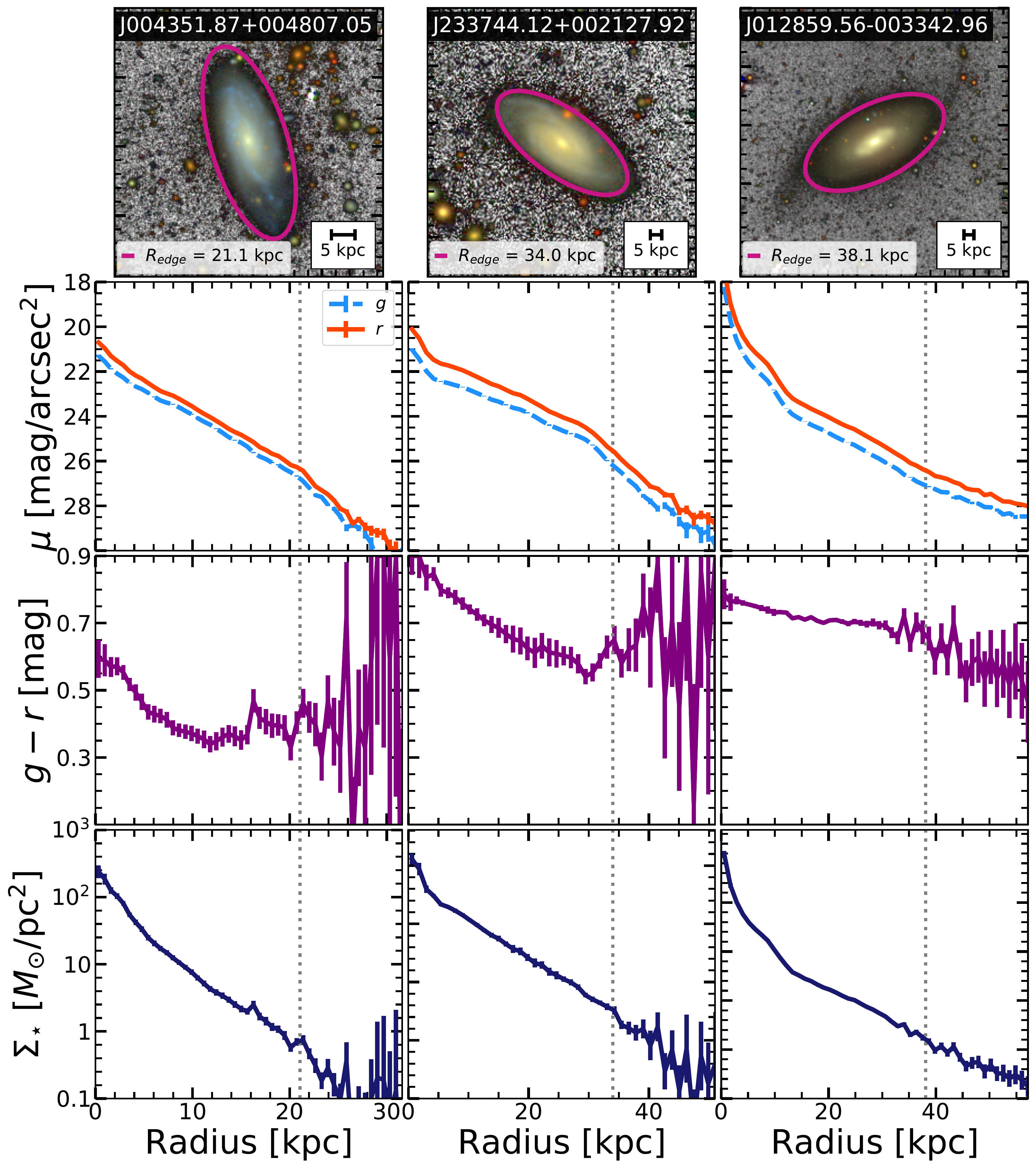}
    \caption{Radial profiles of three difficult cases. The panels are labelled as in Fig. \ref{fig:edges_example}.}
    \label{fig:difficult_cases}
\end{figure*}

Until now, we consider that all the profiles visualised in this work in Figs. \ref{fig:edges_example}, \ref{fig:dwarf_examples} and \ref{fig:elliptical_examples} have edges in their profiles. In this section, we show galaxies where the identification of the edge was less straightforward. We show three difficult cases  in Fig. \ref{fig:difficult_cases} and describe each case below (galaxies ordered from left to right in the figure). These cases represent 10\% of the cases which required more detailed analysis as described in our visualisation procedure (Fig. \ref{fig:flowchart}). The uncertainties quoted in the $R_{\rm edge}$ values were computed following the procedure described in Sect. \ref{sect:methods}. \par 

\begin{itemize}

    \item J004351.87+004807.05: While the edge is clearest in the $\mu_g$ and $\mu_r$ surface brightness profiles, two distinctive features appear in the $g-r$ and $\Sigma_{\star}$ profile: one at $\sim 16$\,kpc and $\sim 21\pm 3$\,kpc. Upon examination of the 2D colour image, it is clear that the galaxy is not perfectly symmetric in the upper half. The first bump in the profile can thus be attributed to the edge in the upper half of the galaxy and the second to the outer edge marked in this work. We prefer the outer feature as it is coincides with the change in slope visible in the surface brightness profiles.
    
    \item J233744.12+002127.92: $\mu_g$ and $\mu_r$ only show the break at $\sim 30$\,kpc. However, the colour and mass density profile consist of an additional feature at 34\,kpc and is marked as the edge here. Beyond this location the colour drops and rises again. The feature marked is preferred over that at 37\,kpc where the colour rises because this does not correspond to a drop in mass density. The estimated uncertainty in the location of $R_{\rm edge}$ for this galaxy is $\pm 5$\,kpc. 
        
    \item J012859.56-003342.96: This galaxy is an example with tidal-like features in its outskirts that break the elliptical symmetry of the main body as mentioned in Sect. \ref{sect:define_edge} and Fig. \ref{fig:flowchart}. The change in slope is clearest in the $g-r$ colour profile where a drop towards bluer colours occur. In such cases, we think that the colour provides the best indication of the change in stellar properties that occur beyond the edge \citep[see also the discussion in][]{2021trujillo}. 
    
\end{itemize}

\section{The effect of image depth}
\label{app:image_depth}

In this appendix, we demonstrate that the criteria we have developed in this work to locate the edges of galaxies (see Fig. \ref{fig:flowchart}) are not biased due to the depth of the IAC Stripe 82 images we used. The analysis presented in this appendix is divided in two parts. In the first, we compare radial profiles at the depth of IAC Stripe 82 with those from the deeper LBT Imaging of Galactic Haloes and Tidal Structures (LIGHTS) Survey \citep[see][for more details]{2021trujillo} using $g$ and $r$-band images of a M33-like spiral galaxy, NGC 1042, as an example case. The images from LIGHTS are publicly available and have limiting depths in surface brightness of 31.2 and 30.5 mag/arcsec$^2$ (3$\sigma$, 10$\times$10 arcsec$^2$) in $g$ and $r$ respectively, with a pixel scale of 0.224 arcsec/pixel. In the second part of this appendix, we examine our capability of identifying the edges of galaxies in our parent sample by comparing the surface brightness at which they appear to the limiting surface brightness of the images IAC Stripe 82 used. \par 

\begin{samepage}

\begin{figure*}
    \centering
    \includegraphics[width=0.8\textwidth]{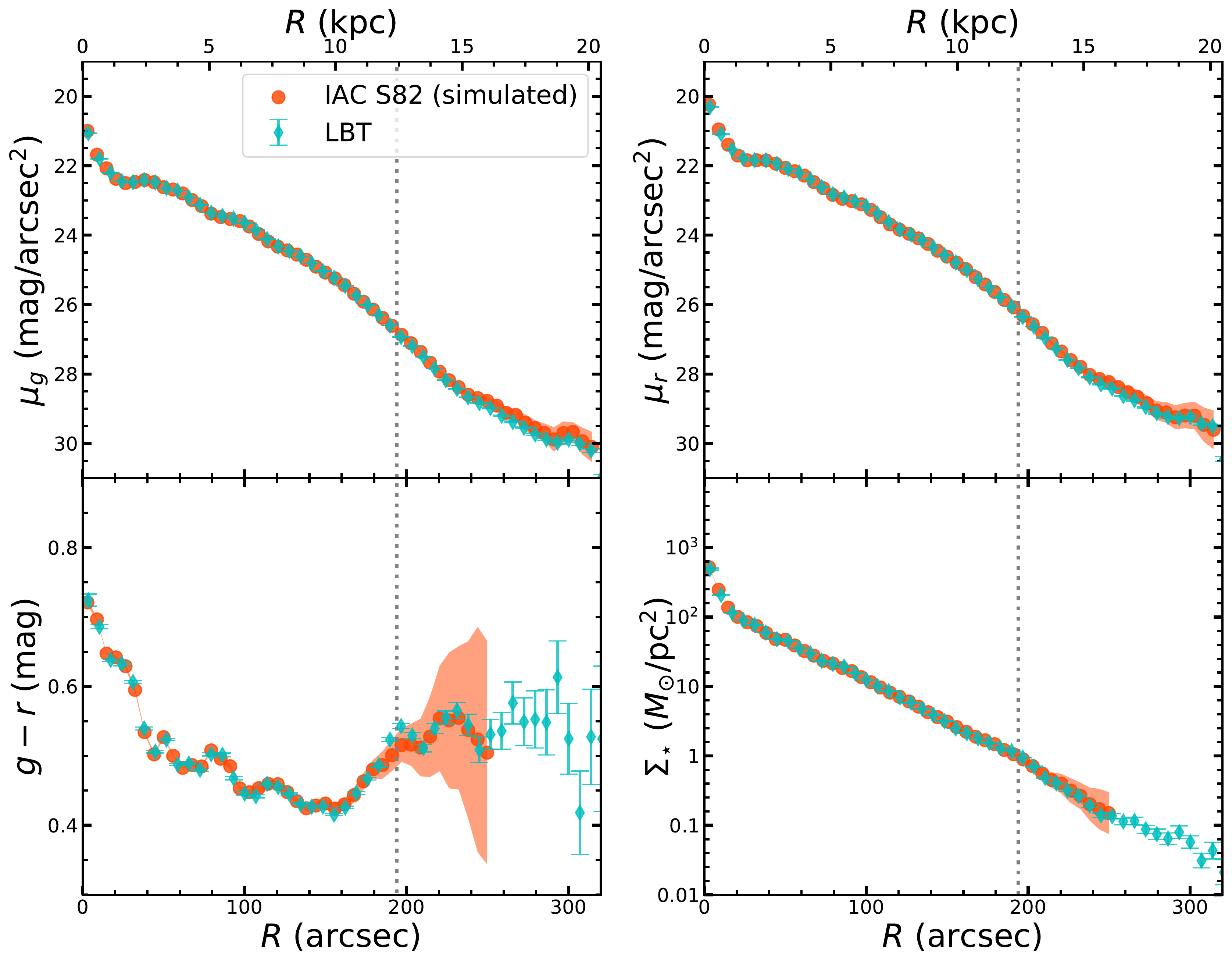}
    \caption{Radial profiles of NGC1042 from LBT \citep{2021trujillo} and simulated IAC Stripe 82 depth (this work, see text for details). The simulated profile is the median of 100 realisations and the shaded regions correspond to $\pm$ three standard deviations up to where the deviation from the median is less than 0.2\,mag. The vertical grey line is located at the edge of the galaxy ($R=194''$). }
    \label{fig:s82_simulated_lbt}
\end{figure*}

\begin{figure*}
    \centering
    \includegraphics[width=1.0\textwidth]{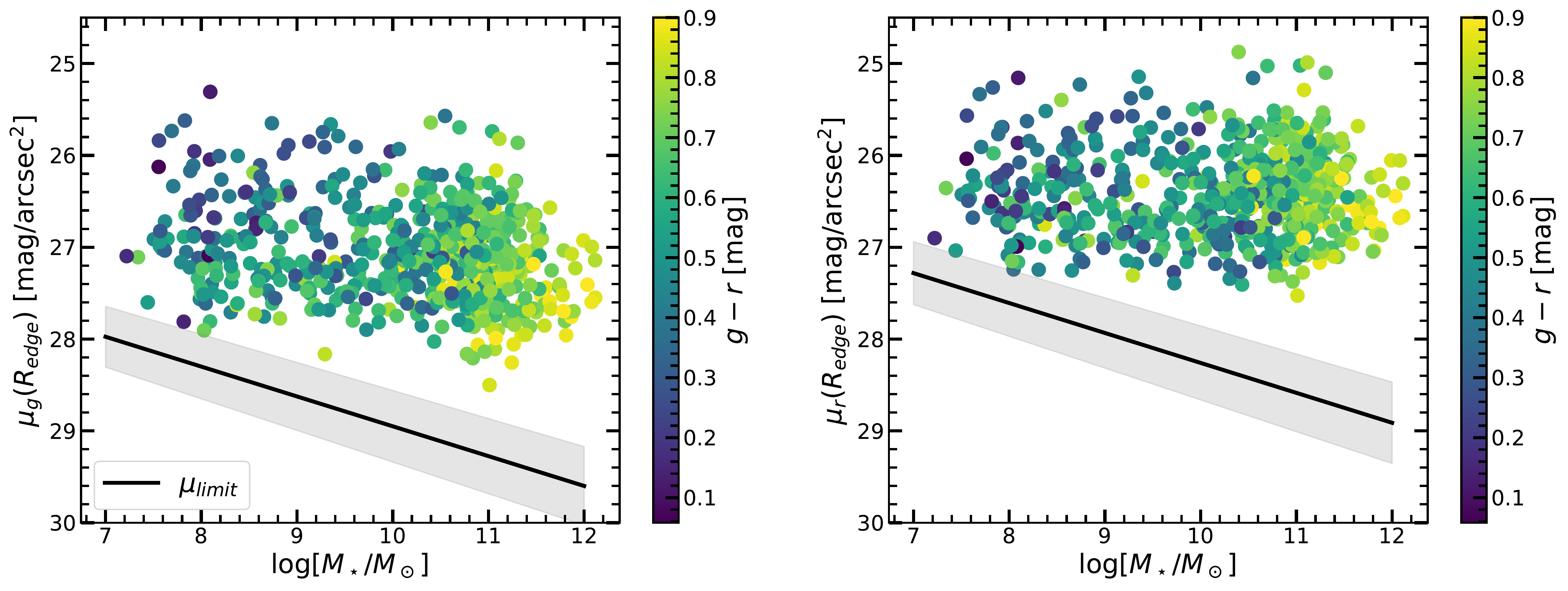}
    \caption{Observed surface brightness in the $g$ (left) and $r$-band (right) at $R_{\rm edge}$ ($\mu_g(R_{\rm edge})$ and $\mu_r(R_{\rm edge})$ respectively) as a function of stellar mass. The edges appear well-within the $3\sigma$ limit of our data computed near $R_{\rm edge}$ (black line, see text for details).}
    \label{fig:mug_mur_edges}
\end{figure*}

\end{samepage}

\subsection{LIGHTS vs. IAC Stripe 82}
\label{app:lbt}

The LIGHTS data of NGC 1042 is interesting to explore for three reasons. Firstly, the LBT images are two magnitudes deeper than IAC Stripe 82 and consequently, compatible with the expected 10-year depth of the future LSST (in the r-band). This allows us to study the edge of a galaxy with a signal-to-noise more than five times higher than with IAC Stripe 82 imaging. Secondly, the edge of NGC1042 has been identified in \citet{2021trujillo} and is one of the deepest observations of an edge to date for a nearby \citep[13.5 Mpc;][]{2019monelli} disk galaxy. Therefore studying the effect of depth in our procedure using these NGC1042 observations allow us to motivate future studies for similar galaxies. Lastly, NGC1042 has similar properties to the galaxy sample studied in P17 which we motivated earlier in Sect. \ref{sect:sample}. In other words, it is a disk galaxy with low inclination (axis ratio of 0.83) and given its distance, allows one to study its structure at a higher spatial resolution (65 pc/arcsec). All of the above reasons are well motivated for the goals of this and future work.  \par 
Unfortunately this galaxy is outside the footprint of Stripe 82. Therefore to study the effect of depth on the edge of NGC1042, we degraded the LIGHTS images to mimic IAC Stripe 82 in the following way. After background subtracting the LBT images \citep[see][]{2021trujillo}, we resampled them to match the SDSS pixel scale (0.396 arcsec/pixel) and zero point (22.5\,mag), and added random gaussian noise to reach IAC Stripe 82 depth (see Sect. \ref{sect:sample}). In other words, we used the typical standard deviation of the background pixels in the LBT ($\sigma_{LBT}$) and IAC Stripe 82 ($\sigma_{S82}$) images based on their limiting depth to compute the required level of noise to degrade the LBT images ($\sigma_{sim}$):

\begin{align}
\sigma_{sim}^2 &= \sigma_{S82}^2 - \sigma_{LBT}^2
\end{align}   

We then followed the procedure outlined in Fig. \ref{fig:flowchart} on the ``simulated'' IAC Stripe 82 images based on the degraded LBT images and locate the edge of the galaxy. For this particular galaxy, we followed the same masking procedure as that outlined in \citet{2021trujillo} to ensure the same regions are compared when interpreting the data. The result is shown in Fig. \ref{fig:s82_simulated_lbt} where we show the radial profiles from LBT (taken from \citet{2021trujillo}) and the median profile with simulated IAC Stripe 82 depth after 100 separate realisations of random noise. The shaded regions correspond to $\pm$ three standard deviations from the median profile plotted up to where the deviation is less than 0.2\,mag.  From the $g-r$ and $\Sigma_{\star}$ profile, we see that the location of the edge ($R = 194''$ marked with the vertical dotted line) does not change between the profiles. The main difference is that the profiles are more noisier in the outskirts with IAC Stripe 82 depth but this does not prevent the identification  of the edge (although it is certainly harder) using our criteria. For NGC1042, the edge appears at a mass density of 1 \,$M_{\odot}$/pc$^2$. \par

\subsection{Detecting edges with IAC Stripe 82 depth}
\label{app:limit_sb}

In the second part of this appendix, we compare the surface brightness at which the edges of our parent sample appear with the limiting surface brightness of the IAC Stripe 82 images used. We do this by computing the representative depth of each image as it depends on the size of galaxies (i.e. the area near $R_{edge}$ for each galaxy). This can be understood in the following way. In general, the limiting surface brightness of an image can be computed as the $x\sigma$ fluctuation (where $x$ corresponds to the number of deviations) with respect to the background of the image that is measured over an area $A$. This is called a metric, i.e. ($x\sigma$;$A$\,arcsec$2$). As a rule of thumb, $A$ is optimally chosen according to the apparent size of objects under consideration (using a circle, box etc.). Therefore, for example in the SDSS g-band, we may write that the limiting surface brightness depth is $\mu_{lim, g} \sim$ 26.5\,mag/arcsec$^2$ (3$\sigma$;R = 12'') , where $A$ was chosen to be the area of a circle with radius R = 12 \citep{2004kniazev}. This is a reasonable choice for $A$ because \citet{2004kniazev} was interested in the search for new dwarf and low surface brightness galaxies in SDSS. To convert the limiting surface brightness (at a given wavelength $\lambda$) from metric ($x_1\sigma$; $A_1$) to ($x_2\sigma$; $A_2$) is:

\begin{equation}
\label{eq:sb_limit}
\mu_{lim,\lambda}(x_2\sigma; A_2) = \mu_{lim,\lambda}(x_1\sigma; A_1) - 2.5\log\left[ \frac{x_2}{x_1} \right ] + 2.5\log\left [ \frac{A_2}{A_1} \right]^{1/2}    
\end{equation}

where $x_k$ is the number of variations $\sigma$ and $A_k$ is the area used for $k = {1, 2}$. Therefore, given that we know the depth of our images for an area of $10\times10$\,arcsec$^2$ (Sect. \ref{sect:sample}), we can use Eq. \ref{eq:sb_limit} to compute the limit over different areas. We do this for the size of each galaxy in our parent sample in the following way to confirm that the depth of the image is enough to detect their edges. \par

We plot the observed surface brightness in the $g$-band at which the edges of the galaxies appear ($\mu_g(R_{\rm edge})$) as a function of stellar mass in Fig. \ref{fig:mug_mur_edges}. To determine the representative limit of the data given the size of the galaxy, we compute the $3\sigma$ limit over the area of pixels in an elliptical annuli (used when deriving the surface brightness profile, see Sect. \ref{sect:methods}) at $R_{\rm edge}$ for each galaxy. The line of best fit is plotted in black and the gray shaded region bounds the $\pm1\sigma$ dispersion of the fit. \par

The figure shows that all the edges studied in this work appear at surface brightnesses above the limit near $R_{\rm edge}$, even for the redder galaxies. Therefore, we may conclude that the edges we study here are not a consequence of image depth and the stratification in the size--stellar mass plane is not an observational bias towards bluer edges. In other words, if there were larger galaxies with redder edges in our sample at any fixed stellar mass, we should have been capable of identifying them with our data.

\section{The dependence of radial profile derivation methods on galaxy orientation}
\label{app:ellipse_semi}

Two methods have been mainly used to derive the radial profiles of galaxies in the literature. In the first method, radial profiles are derived along the galaxy's semi-major axis using a slit of fixed width\footnote{{A slightly different approach to this method is the use of a wedge shape \citep[see][]{2021stone}.}}. The second method uses elliptical annuli: the axis ratio and position angle to parameterise the ellipse for this procedure may either be fixed (this work) or allowed to vary along the main body of the galaxy. The slit method is normally used to characterise the radial profiles of edge-on oriented galaxies as the method is limited to imaging with high signal-to-noise and edge-on galaxies are naturally observed deeper than face-on ones due to the line-of-sight integration. On the other hand, the ellipse method is limited to galaxies that can be described by an ellipse and this is not the case for edge-on or near edge-on ($q \lesssim 0.3$) galaxies with prominent bulges. Therefore, the ellipse method has generally been used to derive the profiles of low-inclination galaxies and is thus possible with lower signal-to-noise imaging. \par 

In this appendix, we study and compare both of these profile derivation methods on an edge-on and low-inclination galaxy to examine its effect on the visibility of the edge and on galaxy orientation. We make use of the LBT data of the low-inclination galaxy NGC1042 as in Appendix \ref{app:lbt} and SDSS DR12 images of the edge-on galaxy UGC09138 shown in Fig. \ref{fig:edges_example}. \par

\begin{figure}
    \centering
    \includegraphics[width=0.5\textwidth]{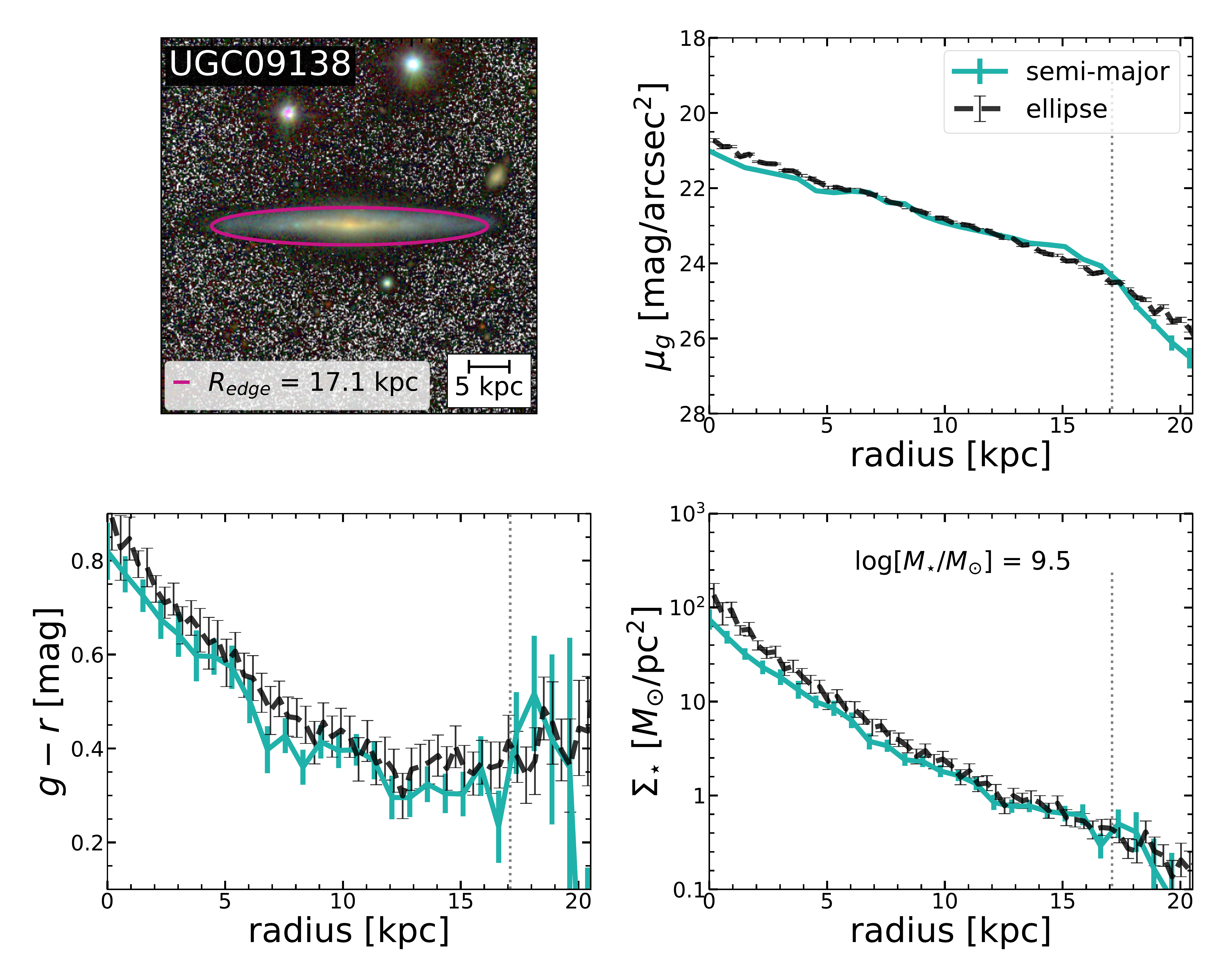}
    \includegraphics[width=0.5\textwidth]{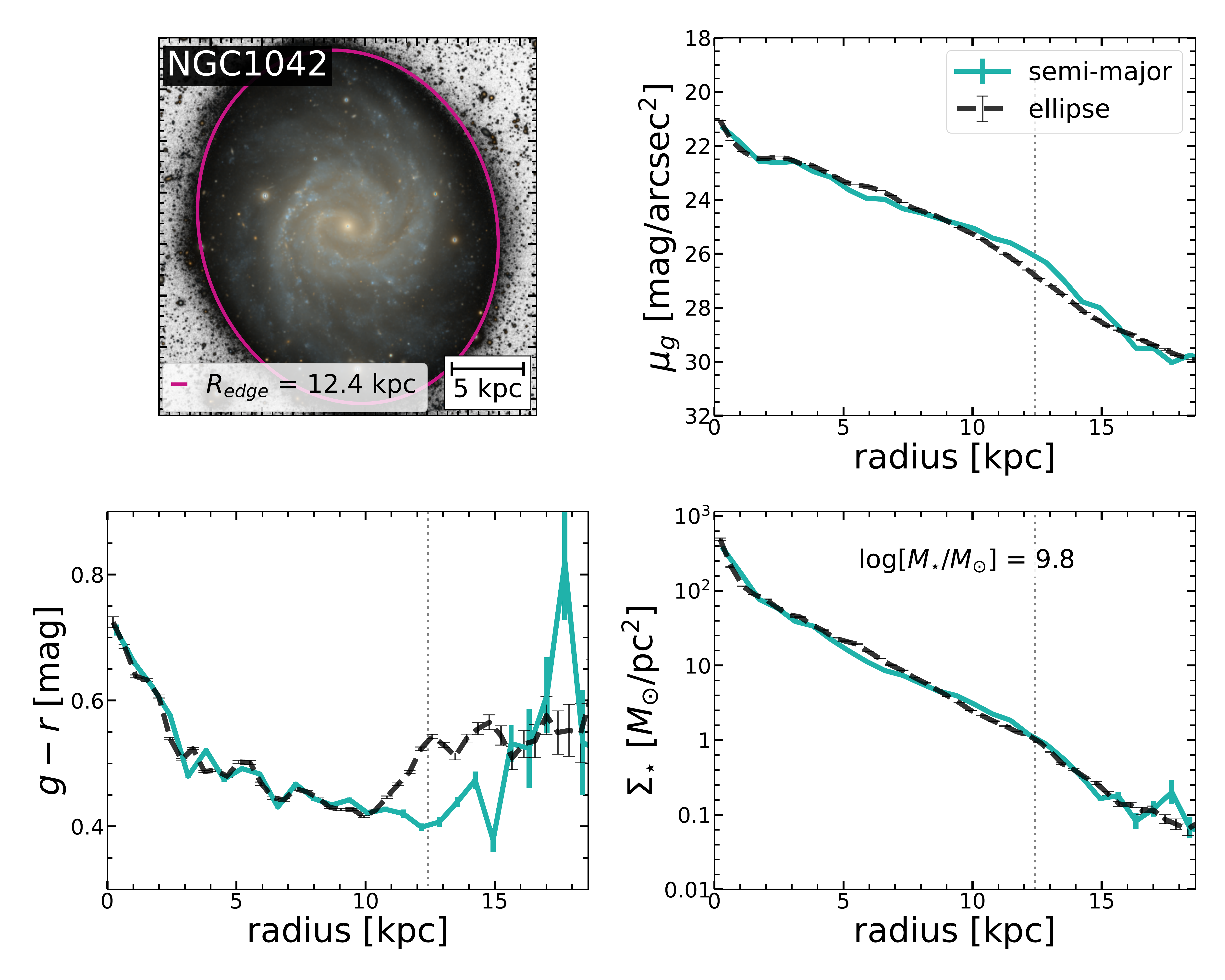}
    \caption{Comparison between radial profiles derived using a 1.2\,kpc slit through the semi-major axis of the galaxy (solid, cyan) and elliptical annuli (dashed, black) for UGC09138 (edge-on, top panels) and NGC1042 (low-inclination, lower panels). The elliptical annuli method is suitable for the low-inclination galaxies studied in this work. See text for details.}
    \label{fig:face_edge_on_galaxy}
\end{figure}

For the slit method, we used a 1.2\,kpc width for both galaxies and we followed the procedure outlined in Sect. \ref{sect:methods} for the elliptical annuli method. The result of the comparison is shown in Fig. \ref{fig:face_edge_on_galaxy} for UGC09138 (top panels) and NGC1042 (lower panels). The profiles derived using the slit method is plotted in solid, cyan lines and ellipse method in dashed, black ones. It can be seen that the edge of UGC09138 may only be located using the slit (semi-major) axis method in the $\mu_g$ and $\Sigma_{\star}$ profiles while both methods may be used to locate the edge of NGC1042 in the $\Sigma_{\star}$ profile. However, the edge also becomes clearer in the $\mu_g$  profile of NGC1042 using the slit method. The latter result thus highlights the fact that deeper imaging with higher signal-to-noise than SDSS or IAC Stripe 82 (e.g. LIGHTS and the future LSST) will allow the identification and characterisation of edges using the slit method also for near face-on ($q \geq 0.8$) galaxies. 

\subsection{Dwarf galaxy SDSS J224114.29-003710.2}
\label{app:d23}

\begin{figure}
    \centering
    \includegraphics[width=0.5\textwidth]{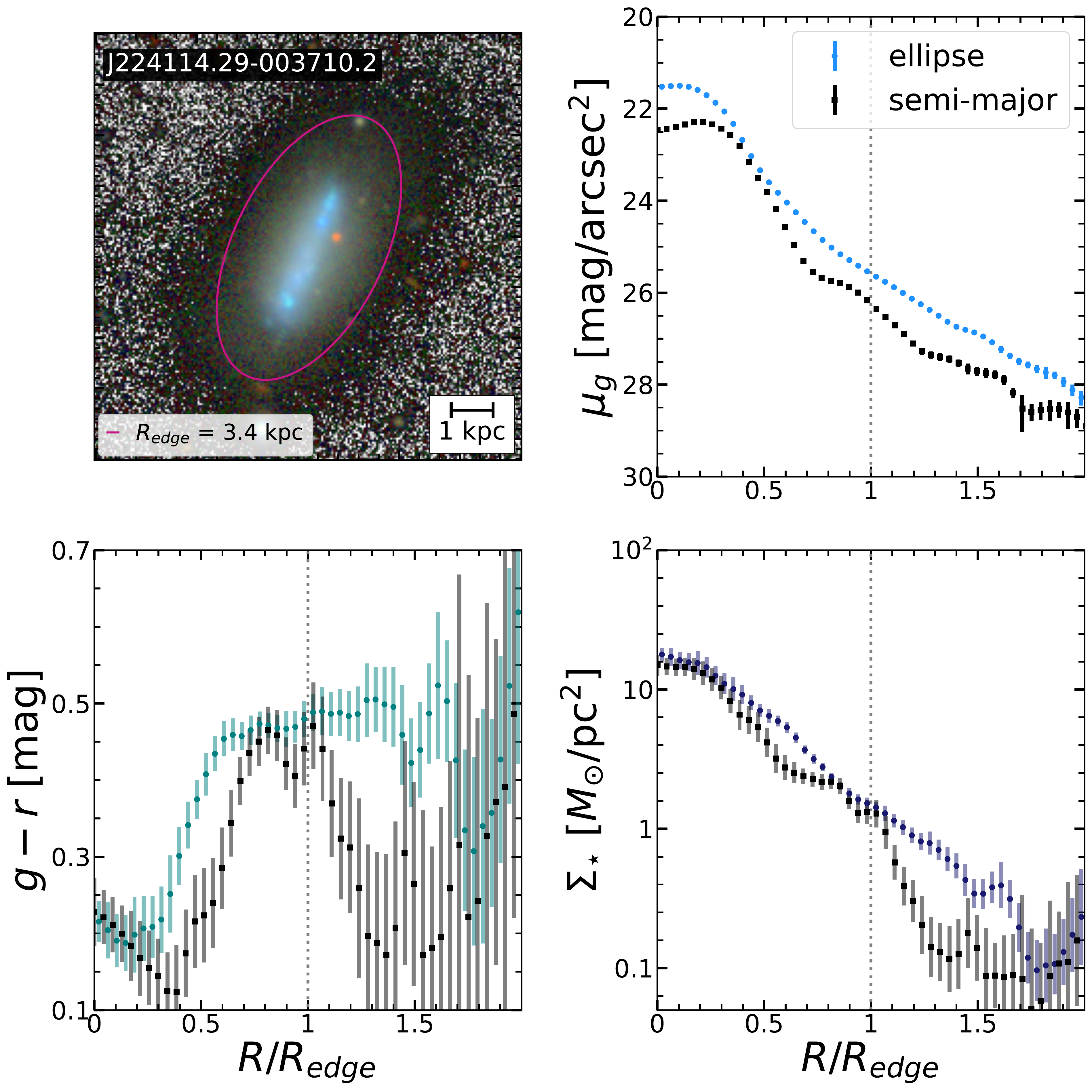}
    \caption{Comparison between radial profiles derived for the dwarf galaxy SDSS J224114.29-003710.2 using a 1\,kpc slit through its semi-major axis (squares, black) and elliptical annuli (dotted, blue). The $x$-axis and $R_{\rm edge}$ for this galaxy are scaled and marked similarly to Fig. \ref{fig:dwarf_examples}.}
    \label{fig:D23_compared}
\end{figure}

We have identified $R_{\rm edge}$ using the change in slope in the stellar mass density profile of the dwarf galaxy SDSS J224114.29-003710.2 in Fig. \ref{fig:dwarf_examples}. The profiles plotted in that figure were derived using the ellipse method as described above and in Sect. \ref{sect:methods}. As an example, we
also derive the profiles using the slit method, with a 1\,kpc width slit through the semi-major axis of the galaxy. The profiles from the two methods are compared in Fig. \ref{fig:D23_compared}. We mark $R_{\rm edge}$ as in Fig. \ref{fig:face_edge_on_galaxy} and confirm that both methods have a signature of the edge in the same location. However, while the slit method additionally provides the signature in the $g-r$ profile as a sharp drop towards bluer outskirts, we point out that this feature does not take into account the global colour of the galaxy at $R_{\rm edge}$ because it was strictly computed within the slit region.

\section{Fits to the $\Sigma_{\star}(R_{\rm edge})-M_{\star}$ relation}
\label{app:fits}

{In this appendix, we justify the use of Eqs. \ref{eq:sigma-edge1} and \ref{eq:sigma-edge2} as best fits to the $\Sigma(R_{\rm edge})-M_{\star}$ relation shown in Fig. \ref{fig:sigma-redge}.}

\begin{figure}
    \centering
    \includegraphics[width=0.5\textwidth]{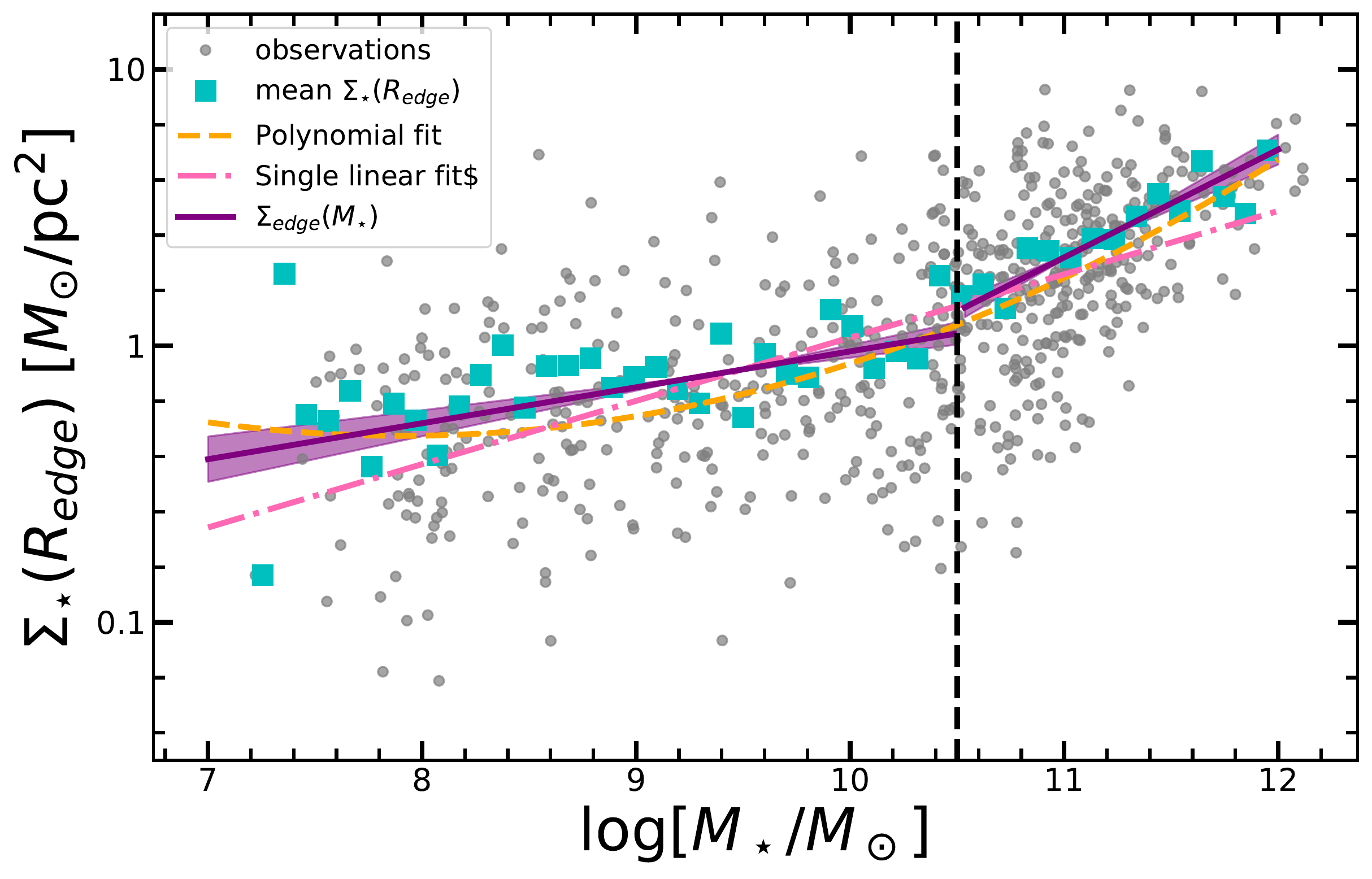}
    \caption{Fitting functions for the $\Sigma_{\star}(R_{\rm edge})-M_{\star}$ relation (grey points). The mean $\Sigma_{\star}(R_{\rm edge})$ is computed in steps of 0.1\,dex over the plotted stellar mass range (cyan squares). The best fits (purple lines) are the two linear relations in Eqs. \ref{eq:sigma-edge1} and \ref{eq:sigma-edge2}, split at $10^{10.5}\,M_{\odot}$ (vertical black dashed line). For contrast, we also show a single linear fit (pink dot-dashed line) and polynomial fit of degree two for the full sample (orange dashed line).}
    \label{fig:fits_sigma}
\end{figure}

In Fig. \ref{fig:fits_sigma}, we compare fitting the measurements using two linear fits (as adopted in this work) with a single linear relation and a polynomial of degree 2. We  test whether these fitted relations pass through the mean $\Sigma_{\star}(R_{\rm edge})$ in steps of 0.1 dex over our full stellar mass range ($10^7\,M_{\odot} < M_{\star} < 10^{12}\,M_{\odot}$). The grey points show our measurements (Fig. \ref{fig:sigma-redge}), the cyan squares show the mean $\Sigma_{\star}(R_{\rm edge})$ in each stellar mass bin, the pink dot-dashed line shows the single linear fit, the orange dashed line shows the  polynomial fit of degree 2 and the purple lines show the linear relationships split at the stellar mass of $10^{10.5}\,M_{\odot}$ (Eqs. \ref{eq:sigma-edge1} and \ref{eq:sigma-edge2}). We mark where the relations split with a vertical dashed black line. The shaded region shows the 1 sigma scatter in the fitted linear relations.\par
\begin{figure}
    \centering
    \includegraphics[width=0.5\textwidth]{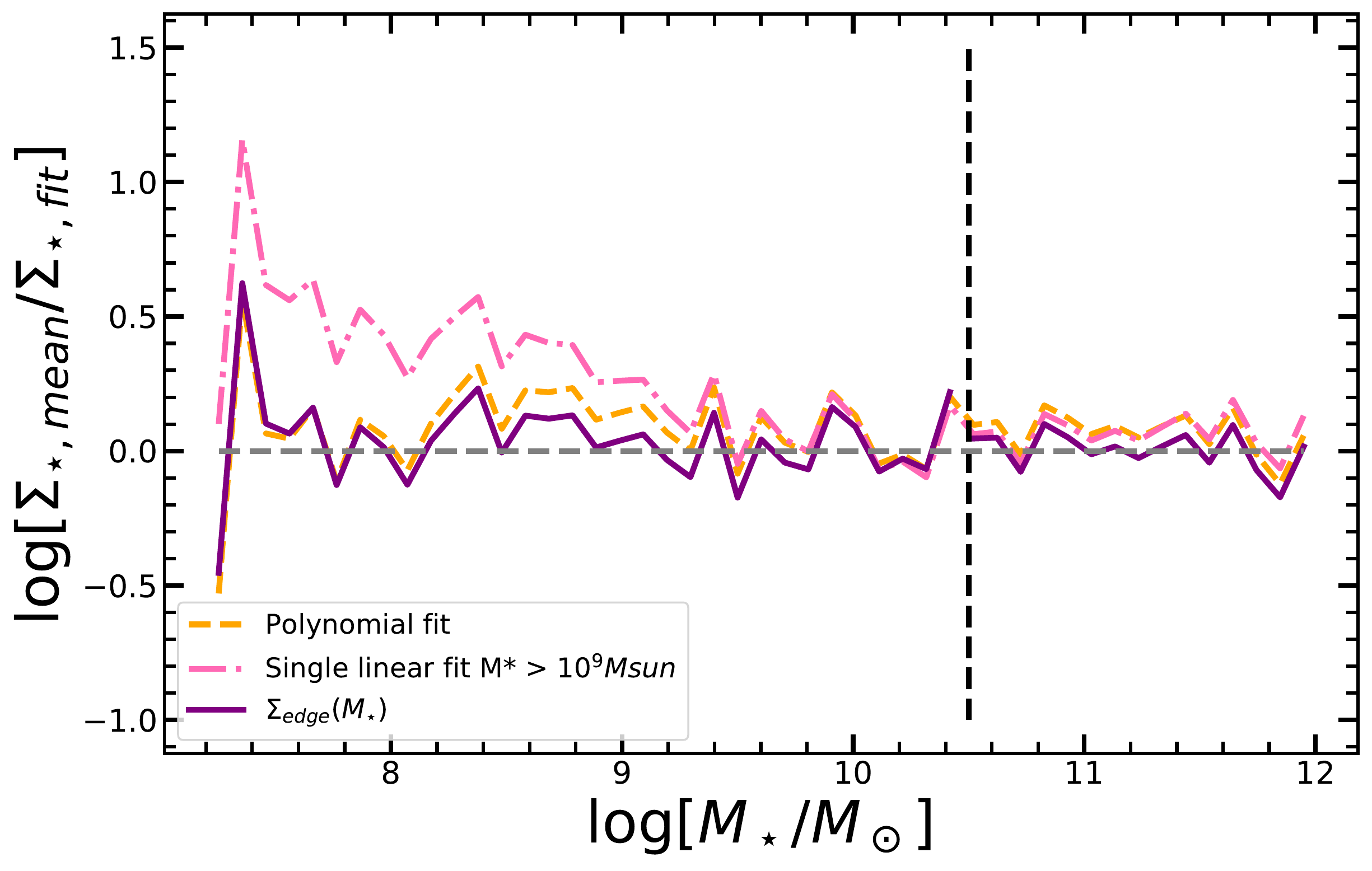}
    \caption{Comparing the residuals of the fits to the mean $\Sigma_{\star}(R_{\rm edge})$ relation. The single linear fit in pink used here is computed for galaxies with $M_{\star} > 10^9\,M_{\odot}$. The reduced chi-square values are 1.6 and 1.7 for the purple $\Sigma_{\rm edge}(M_{\star})$ relations in Eqs. \ref{eq:sigma-edge1} and \ref{eq:sigma-edge2}, 3.1 for the pink linear relation and 2.9 for the polynomial fit.}
    \label{fig:fits_residual}
\end{figure}

 
From this figure, it is clear that both the single linear and polynomial fits do not pass through all the mean points (cyan) of the $\Sigma_{\star}(R_{\rm edge})$–stellar mass plane. The relation also changes slope at a stellar mass of $10^{10.5}\,M_{\odot}$ (vertical dashed line). These findings do not change if we restrict the single linear fit to galaxies only with stellar masses $M_{\star} > 10^9\,M_{\odot}$, which we confirm by studying the residuals of the fits with respect to the mean data points shown in Fig. \ref{fig:fits_residual}. The reduced chi-square values are 1.6 and 1.7 for the purple $\Sigma_{\rm edge}(M_{\star})$ relations in Eqs. \ref{eq:sigma-edge1} and \ref{eq:sigma-edge2}, 3.1 for the pink linear relation and 2.9 for the polynomial fit, with respect to the mean data points.

We also confirm that these results do not change if we exclude the dwarf galaxies in our sample. Therefore, we use the two linear relations as the best fits to our measurements in the $\Sigma_{\star}(R_{\rm edge})$–stellar mass plane. This choice also leads to comparable sample sizes of galaxies with stellar masses either greater or less than $10^{10.5}\,M_{\odot}$. \par 
Finally, for the late-type galaxies plotted in Fig. \ref{fig:late_types}, we point out that while the scatter of the relation is large, the Spearman’s correlation coefficient is positive and with a p-value $<$ 1\%. This result is an indication that there is a very strong relation in this parameter space even though the scatter is high.

\section{The usability of $R_{\rm edge}$ in large-scale, multi-band surveys}
\label{sect:large_cats}

In this work, we have visually identified the edges of a large sample of galaxies (645 galaxies). Such a detailed study will no longer be feasible once data will be acquired from wide surveys such as the LSST mentioned above, which is expected to provide 20 TB of data per night\footnote{\protect\url{https://www.lsst.org/scientists/keynumbers}}. Therefore, a natural question is how convenient and usable is $R_{\rm edge}$ as a size measure to be included in large-scale galaxy catalogues? \par
\begin{figure}
     \centering
     \includegraphics[width=0.5\textwidth]{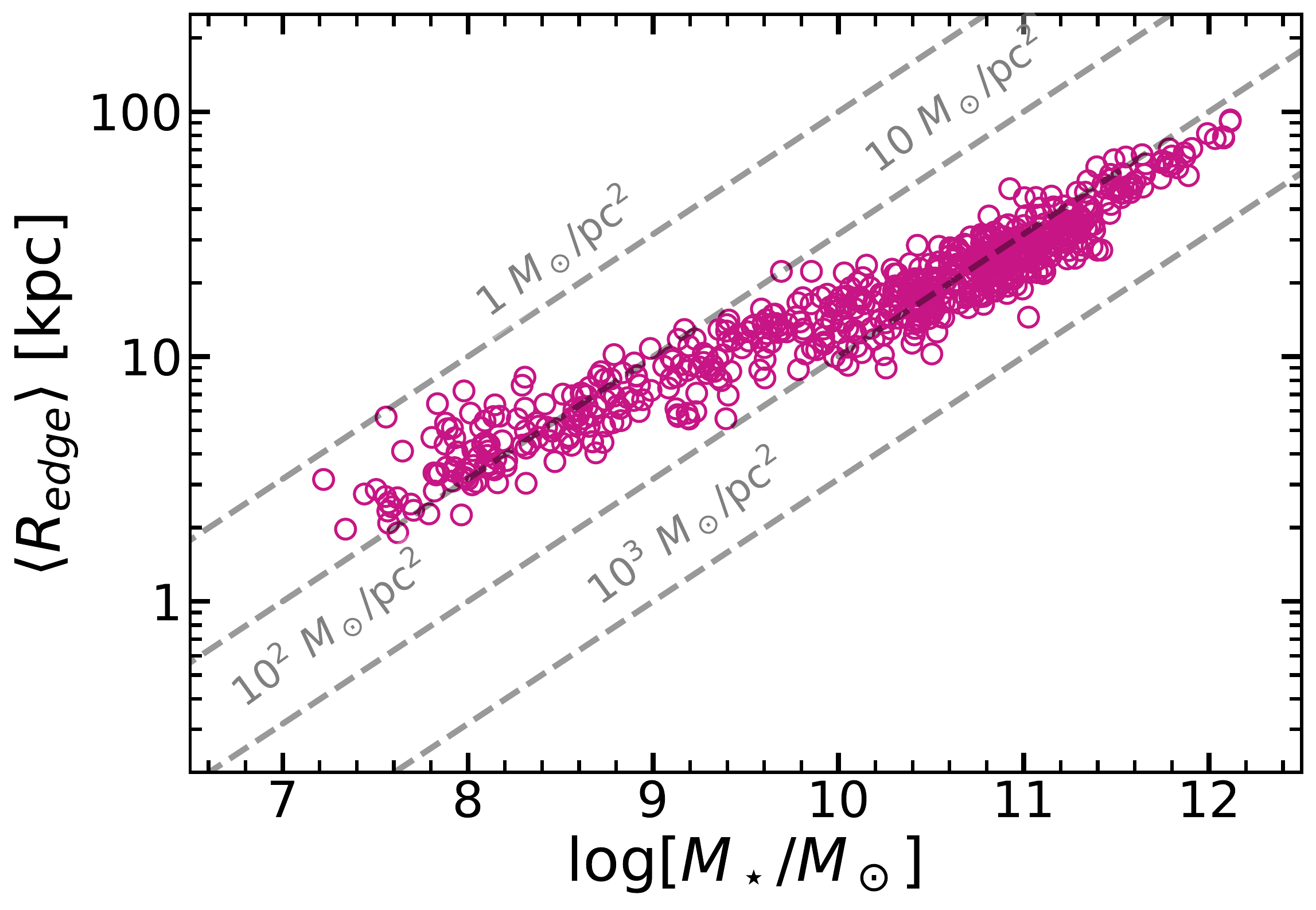}
     \caption{$\langle R_{\rm edge} \rangle$--stellar mass relation for the parent sample resulting from the density threshold laws in Eqs. \ref{eq:sigma-edge1} and \ref{eq:sigma-edge2}. The structure of the relation is almost exactly the same as the observed $R_{\rm edge}$--stellar mass relation shown in Fig. \ref{fig:sigma-redge} and is therefore a promising method to determine a proximal measure of $R_{\rm edge}$ automatically for large-scale cataloguing.}
     \label{fig:redge_mass_fitted}
 \end{figure}

 
We emphasize that we use the stellar mass density at the location of the edge as a \textit{proxy} for the underlying theoretical gas density threshold required for star formation in galaxies \citep[see][]{2004schaye, 2020tck}. For this reason, as an exercise, we used the $\langle \Sigma_{\rm edge}(M_{\star})\rangle$ relations (Eqs. \ref{eq:sigma-edge1} and \ref{eq:sigma-edge2}) to ascertain the average location of the edge for each galaxy, $\langle R_{\rm edge} \rangle$. For each galaxy with stellar mass $M_{\star}$ we locate the position of $\langle \Sigma_{\star}(R_{\rm edge}) \rangle$  using the stellar mass density profile of the object, i.e. $\Sigma_{\star}(\langle R_{\rm edge}\rangle)  =
 \langle \Sigma_{\rm edge}(M_{\star})\rangle$. The resulting $\langle R_{\rm edge} \rangle$--stellar mass relation for the parent sample is shown in Fig. \ref{fig:redge_mass_fitted}. The $\langle R_{\rm edge} \rangle$--stellar mass plane also follows a power law of the form $\langle R_{\rm edge} \rangle \propto M_{\star}^{\langle \beta \rangle}$. As expected, the dispersion of this plane is less than that of the observed $R_{\rm edge}$--stellar mass plane: $\sigma_{\langle R_{\rm edge} \rangle} = 0.093 \pm 0.005$ dex but the slope is compatible $\langle \beta \rangle = 0.30 \pm 0.006$. 
 
The above result suggests that Eqs. \ref{eq:sigma-edge1} and \ref{eq:sigma-edge2} could potentially be used to obtain $\langle R_{\rm edge} \rangle$ and provide a proxy for the size of any galaxy, provided its stellar mass is known. These laws can be useful for larger galaxy samples and automated cataloguing in future multi-band surveys such as the LSST. Providing $\langle R_{\rm edge} \rangle$ is also advantageous for cases where the edge is not apparent and thus serves as a prediction of where the edge should be for a given galaxy. Strong deviations from this prediction could then provide insights about the stellar population properties in the outskirts of the galaxy in comparison with the parent population. \par 
We point out that galaxies with low stellar density and mass (i.e. low surface brightness galaxies, e.g. \citet{1984sandage}) are not included in our sample and therefore the validity of the laws presented in this work to such galaxies need to be investigated in future work \citep[but see][]{2020ctk}. 

\end{document}